\documentclass[12pt]{article}
\pdfoutput=1
\usepackage[english]{babel}
\usepackage{graphicx}
\usepackage[T1]{fontenc}
\usepackage[utf8]{inputenc}
\usepackage{lmodern}
\usepackage{bbm}
\usepackage{color}
\usepackage{amsmath}
\usepackage{cite}
\usepackage{url}

\def\be{\begin{eqnarray}}
\def\ee{\end{eqnarray}}

\newcommand{\Tr}{\mbox{Tr}}

\newcommand{\eq}{\begin{equation}}
\newcommand{\eqx}{\end{equation}}
\newcommand{\eqn}{\begin{eqnarray}}
\newcommand{\eqnx}{\end{eqnarray}}
\newcommand{\ben}{\begin{eqnaray}}
\newcommand{\een}{\end{eqnarray}}

\newcommand{\GG}{{\cal G}}
\newcommand{\cG}{\GG}
\newcommand{\cX}{{\cal X}}
\newcommand{\cY}{{\cal Y}}
\newcommand{\cQ}{{\cal Q}}
\newcommand{\cA}{{\cal A}}

\newcommand{\ZZ}{{\cal Z}}

\newcommand{\HH}{{\cal H}}
\newcommand{\RR}{{\cal R}}

\newcommand{\zb}{\bar{z}}

\newcommand{\arr}[4]{
\left(\begin{array}{cc}
#1&#2\\
#3&#4
\end{array}\right)
}

\newcommand{\btr}{{\rm bTr}}
\newcommand{\One}{\mbox{\bf 1}}

\newcommand{\lm}{\lambda}

\newcommand{\ket}[1]{\left| #1 \>}
\newcommand{\bra}[1]{\< #1 \right|}
\newcommand{\braket}[2]{\< #1 | #2 \>}

\newcommand{\bb}[1]{{\bf #1}}
\newcommand{\<}{\left<}
\renewcommand{\>}{\right>}
\newcommand{\wb}{\bar{w}}
\newcommand{\idm}{\One}
\newcommand{\vb}{\bar{v}}
\newcommand{\gb}{\bar{g}}

\begin{document}


\title{Spectra of large time-lagged correlation  matrices from Random Matrix Theory}

\author{
 Maciej
A. Nowak\footnote{e-mail: {\tt maciej.a.nowak@uj.edu.pl}}\,\,  and  
Wojciech Tarnowski\footnote{e-mail {\tt wojciech.tarnowski@uj.edu.pl}}\\
 \small M. Smoluchowski Institute  of Physics and \\
\small Mark Kac Complex Systems Research Center,
\small Jagiellonian University,\\
\small S. \L{}ojasiewicza 11,\\
\small PL 30-348 Krak\'{o}w, Poland.}

\date{\today}

\maketitle

\begin{abstract}
We analyze the spectral properties  of large, time-lagged correlation matrices using the tools of random matrix theory. 
We compare predictions of the one-dimensional spectra, based on approaches already proposed in the literature.
Employing the methods  of free random variables and diagrammatic techniques, we solve a general random matrix problem, namely the spectrum of a matrix $\frac{1}{T}XAX^{\dagger}$, where $X$ is an $N\times T$ Gaussian random matrix and $A$ is \textit{any} $T\times T$, not necessarily symmetric (Hermitian) matrix.  Using this result,  we study the spectral features of the large lagged correlation matrices as a function of the depth of the time-lag.  We also analyze the properties of left and right eigenvector correlations  for the time-lagged matrices.   We positively verify our results by the numerical simulations. 

\end{abstract}




\pagebreak

\tableofcontents

\section{Introduction}

Finding the casual relationship  among  several  stochastic  series of signals corresponding to $N$ sources represents a formidable challenge. The role of the statistical analysis  for this task was already noticed by Masani and Wiener in the sixties of the XX century~\cite{WIENERMASANI}, to be followed by Granger~\cite{GRANGER}  and developed  by many~\cite{MANY,MANY2,MANY3,MANY4,MANY5,MANY6}.  In the multivariate analysis, the study of the cross-correlations is  perhaps  the most common method. Historically, Wishart~\cite{WISHART} was the first to ask  how to generalize the chi-squared distribution for the case of multiple dimensions, corresponding to the matrix of Pearson correlation coefficients of purely random time series. 
Wishart ensemble may be also the first random matrix application in science. If we consider multivariate time series represented by a matrix $X_{it}$, where the 'space' index takes values from $\{1, \dots, N\}$, 
and 'time' index  $t\in\{1, \dots, T\}$, the empirical correlation matrix  can be represented as $C_{ij}=\frac{1}{T} \sum_{t=1}^T x_{it}\bar{x}_{jt}$ or, in a matrix notation,  
$\bb{C}=\frac{1}{T}\bb{xx}^{\dagger}$, where  bar ($\dagger$, respectively) applies to the more general case  of the data valued in complex numbers. Lowercase $x$ denotes  the standardized time series, i.e. after a procedure  when for each individual time series we subtract the corresponding mean and divide the result be the corresponding variance.
Wishart ensemble corresponds to a maximally random correlation matrix, where 
each entry $x_{it}$ of a rectangular signal matrix is drawn from a real (complex) Gaussian distribution, so 
$\<x_{it} \bar{x}_{jt^{'}}\>=\delta_{ij}\delta_{tt^{'}}$, where angle brackets  represent taking the expectation values with respects to the probabilistic measure (here Gaussian).  In general, the time series may include
correlations. In particular, assuming space and time factorization,  $\<x_{it} \bar{x}_{jt^{'}}\>=A_{ij}B_{tt^{'}}$ where $\bb{A}$ and $\bb{B}$ are symmetric (Hermitian) positive definite matrices. The spectral properties of such  single-correlated  ($\bb{A}=\idm_N $ or $\bb{B}=\idm_T$) \cite{WISHSINGLE,WISHSINGLE1,WISHSINGLE2,WISHSINGLE3} or doubly-correlated ($\bb{A} \neq \idm_N $ and  $\bb{B} \neq \idm_T$) \cite{WISHDOUBLY,BURDAJURKIEWICZ}
Wishart ensembles were extensively studied in the literature. 

When the number of consecutive measurements (the length of the series) tends to infinity the empirical correlation matrix tends to the true correlation matrix. Unfortunately, there are systems in which one cannot repeat measurements due to uniqueness of data (e.g. study of climate) or the fact that the number of independent time series is of the same order as the number of measurements (e.g. stock markets). In such instances the empirical correlation matrix deviates from the true correlation matrix. The Wishart matrix corresponding to such situation can be obtained in a limiting procedure $N,T\to\infty$ with $r=N/T$ fixed. This limit is referred to as Random Matrix Theory (RMT) limit or Big Data limit~\cite{BOUCHAUDREVIEW}. The parameter $r$, pertaining to the rectangularity of the array of collected data, is also known as the signal-to-noise ratio. Even having the ability to collect a vast number of data points, one would like to extract the leading portion of information from the system. Such a procedure relies on the diagonalization of the correlation matrix and focusing on the largest eigenvalues and corresponding eigenvectors and bears the name of Principal Component Analysis (PCA).

The Wishart ensemble has found applications in telecommunication~\cite{TELECOM} and quantum information~\cite{QUANTUM}.
The applications of PCA in the Big Data limit range from financial  engineering ~\cite{FIN1,FIN2}, through  genetics~\cite{GENETICS,GENETICS2}, meteorology and oceanography~\cite{METEOROLOGY}, study of atmosphere~\cite{ATMO}, climate change detection~\cite{CLIMATE}, criminal offence records~\cite{CRIMINAL}, to the analysis of EEG data~\cite{EEG}. 

The analysis of correlation matrices presented above gives an insight into the interdependence of the components of complex systems \textit{at} the same moments. In order to gain the access to the correlations at different instants, one has to shift the time index when correlating the time series, obtaining the auto cross-correlation matrix $C_{ij}^{\tau}=\frac{1}{T-\tau}\sum_{t=1}^{T-\tau}x_{it}\bar{x}_{j,t+\tau}$, which explicitly depends on the time lag $\tau$. Such object appeared in the literature also under the name of delayed or time-lagged correlation matrix. For simplicity, we refer to this as the lagged correlation matrix. By construction $\bb{C}^{\tau}$ is not symmetric, which reflects the asymmetric influence of the component of the system. Such an information is not accessible by means of equal time correlation matrices.

The analysis of time-lagged correlations is recently becoming an area of a rapid development~\cite{LAG1,LAG2,LIVAN,LAG3,LAG4,LAG5}. The diagonalization of the asymmetric lagged correlation matrices in the financial context was proposed in \cite{DROZDZ} and in studies of human brain activity in \cite{DROZDZ2}. Later on, Podobnik et al. studied~\cite{PODOBNIK} the singular values of the lagged correlation matrices in various complex systems, including EEG signal and stock market, finding the long-range magnitude correlations.

Because of the asymmetry of the lagged correlation matrix, its spectrum is complex, which invalidates the standard tools of random matrix theories. Finding the RMT benchmark of the spectrum, corresponding to the absence of any correlations within the system is a formidable challenge. Due to the importance of the lagged correlations, this spectral problem was attacked few times in the literature, using the tools of RMT: \\
(i) \textit{Symmetrization.} First, the symmetrized  lagged covariance matrix $\bb{C}^{{\rm sym}}=\frac{1}{2}\left(\bb{C}^{\tau}+(\bb{C}^{\tau})^{\dagger}\right)$ was studied~\cite{HINDU,USREVIEW}.  In the Big Data limit ($N,T \rightarrow \infty$ with $N/T$ fixed) the  resulting real spectral density comes from the resolvent (Green's function) fulfilling a quartic algebraic  equation (Ferrari equation).\\
 (ii) \textit{Spectral whitening.} The singular value decomposition of the product $(\bb{C}^{\tau})^{\dagger} \bb{C}^{\tau}$ was studied, after removing all equal-time correlations in order to define maximally random product~\cite{BOUCHAUD}. In the RMT limit, the resulting spectrum is  given by the so-called free Jacobi measure.\\
(iii) \textit{Abelization.} Third,  the strictly non-Hermitian lagged correlation matrix problem was approached in ~\cite{THURNERBIELY}, using  the inverse Abel transform for a circularly symmetric spectrum.   In this case, the radial spectral function stems from the quartic algebraic equation from the approach (i). We call this method 'Abelization', as it refers to the Abel transform and works only for matrices that commute with their Hermitian conjugate.\\
(iv) Finally, the explicit non-Hermitian lagged correlation matrices were analized by Jarosz~\cite{JAROSZLAG} using the non-Hermitian diagrammatic techniques. In particular, the result (iii) was challenged -  the radial spectral  function  was originating from the cubic algebraic equation (Cardano type).\\ 
(v) Later on, Livan and Rebecchi~\cite{LIVAN} proposed another approach relying on the assumption that the benchmark for the lagged correlation matrix can be approximated by a product of two independent rectangular Gaussian matrices, the spectral density of which was calculated in~\cite{BJLNSproducts} and is given by the solution of a quadratic equation.

In this paper, we first summarize the above approaches using the powerful version of Random Matrix Theory in the large size limit, called  free random variables. 
This technique represents a remarkable short-cut for calculations of various correlation matrices, in the limit when the number of measured components $N$ and the number of measurements $T$ are large, but their ratio $r=N/T$ is finite~\cite{USREVIEW,BOUCHAUDLACES}.
Section~\ref{sec:FRV} summarizes the basic operational tools of this technique. Section~\ref{sec:FRV2}  shows its applications in the approaches (i)-(iv).  In particular, we resolve the controversy between~\cite{THURNERBIELY} and ~\cite{JAROSZLAG}, in favor of~\cite{JAROSZLAG}. This is the first  main result of  this paper. 

In section~\ref{sec:diagrammatic}, we recall a certain linearization method for study the non-Hermitian random matrix ensembles. We present the diagrammatic construction in the asymptotic limit. To find the spectrum of the lagged correlation matrix, we solve a more general problem, namely the spectrum of the matrix $\bb{Y}=\frac{1}{T}\bb{XAX}^{\dagger}$, where $\bb{X}$ is an $N\times T$ Gaussian random matrix and $\bb{A}$ is \textit{any} matrix, indepenent of $\bb{X}$, not necessarily symmetric (Hermitian). As a next new result, we also present the  left-  right-eigenvectors correlations for such a problem. To the best of our knowledge, such analysis has never been done before.  
Then, in section~\ref{sec:Analysis}, we apply this result to the particular instance of the lagged correlation matrices also in a double scaling limit $N,T,\tau\to\infty$ with $r=N/T$ {\it and} $\tau/T$ fixed and compare  with the numerical simulation.  Section~\ref{sec:Conclusions} concludes the paper and outlines the further prospects of this method.  To facilitate the reading of the manuscript, we relegated several technical details and the thorough analysis of Feynman diagrams to the appendices. 

\textit{Notation}. In order to avoid ambiguity we adopt a notation at which matrices of a standard size are written with a boldface font while its elements in a standard way. In the sections devoted to the non-Hermitian matrices we denote the matrices with a doubled block structure by the calligraphic letters.


\section{Addition and multiplication theorems for random ensembles} \label{sec:FRV}
\subsection{Real spectra}
In the classical probability theory, the problem of finding the probability distribution function (pdf) $p_{1+2}(s) $    of a sum $s$ of two independent random variables $x$ and $y$ resulting from the corresponding  measures  $p_1(x)dx$ and $p_2(y)dy$ is solved with the help of the characteristic function.  The characteristic function, which is in fact the Fourier  transform of the pdf, factorizes the convolution problem $p(s)=\int dx dy p_1(x)p_2(y) \delta(s-(x+y))$ into a product of the individual characteristic functions for both pdfs.  Since the characteristic function is the generating function for the moments of $p_{1+2}(x)$,  the calculation of moments of the sum boils down to combinatorics. Further simplification comes from considering the logarithm of the characteristic function - the multiplication of the characteristic functions is then replaced   by the addition of their logarithms. This sum rule  holds for each  term in the series expansion,   so the corresponding coefficients, called cumulants, are additive under convolution.  
Similarly, the Mellin transform  reduces the problem of finding probability distribution for the product $z=xy$ of the aforementioned variables to the simple multiplication of the corresponding problems~\cite{EPSTEIN}.

Free random variable (FRV) calculus provides tools, which superimpose the above addition and multiplication laws in the case, when variables $x$ and $y$ are replaced by  asymptotically infinite, non-commuting matrices $\bb{X}$ and $\bb{Y}$. One asks what the spectral measure of $\bb{X}+\bb{Y}$ and $\bb{XY}$ is, provided that the individual spectral measures for both $\bb{X}$ and $\bb{Y}$ are known. As such, FRV represents a certain non-commutative counterpart  of a classical probability calculus, ideally suited for the multivariate analysis of large matrices.  
The main objects in FRV are defined as follows.
We are interested in finding the distribution of the {\it real} eigenvalues
$\lambda_i $ of some stochastic matrix $\bb{H}$  in the limit when  $N$ (size of the matrix) tends to infinity.  It is convenient to introduce the {\it complex} traced resolvent (Green's function)
\be G_\bb{H}(z)=\frac{1}{N} \left\langle {\rm Tr}\, 
\frac{1}{
z\idm_N-\bb{H}}\right\rangle \,. \label{green} 
\ee 
where the brackets $\<...\>$ represent the averaging over the ensemble of $N \times
N$ random Hermitian matrices generated from the probability distribution function
 \be P(\bb{H})
\propto e^{-N {\rm Tr} V(\bb{H})} . \label{probab} \ee 
 On the basis of the Sochocki-Plemelj formula 
 \begin{equation}
 \lim_{\epsilon\to 0} \frac{1}{x \pm i \epsilon} = P.V. \frac{1}{x} \mp i  \pi \delta (x),
 \end{equation}
  the imaginary part of $G(z)$  reconstructs the desired spectral density, 
\be
-\frac{1}{\pi} \lim_{\epsilon \rightarrow 0^+} G(z)|_{z=\lambda+i\epsilon}=\< \frac{1}{N} \sum\limits_{i=1}^{N} \delta (\lambda -\lambda_i) \> \equiv  \rho(\lambda).
\ee
The expansion of the  Green's function for large $z$ generates all spectral moments  $\mu_n=\frac{1}{N} \left \langle {\rm Tr} \bb{H}^n \right \rangle$, alike the characteristic function generates the moments of the pdf in classical probability calculus
\begin{equation}
G(z)=\sum_{k=1}^{\infty}\frac{1}{z^{k+1}} \<\frac{1}{N}\Tr \bb{H}^k\>. \label{eq:HermMoments}
\end{equation}
  Motivated by the concepts of the classical probability calculus, Voiculescu~\cite{VOICULESCU} introduced the 
R-transform as the generating function for cumulants
\be
R(z)=  
\sum_{n=1}^{\infty} \kappa_n z^{n-1},
\label{Rdef}
\ee
i.e. $R(z)$ is the analogue of the logarithm of the characteristic function in probability theory. 
Both complex functions $G(z)$ and $R(z)$ are related via
\be
G\left[R(z)+\frac{1}{z}\right]=R[G(z)] +\frac{1}{G(z)}=z\, .
\label{Rtransform} 
\ee
As an example we consider the instance, which is an analogue of the centered Gaussian, i.e. its distribution is completely determined by the second cumulant only, since all other vanish. Its FRV analogue is therefore $R_s(z)=\kappa_2 z$, which by (\ref{Rtransform}) leads to the quadratic equation for the Green's function with solutions
\be
G_s(z)=\frac{1}{2\kappa_2 z} \left(z \mp \sqrt{z^2-4 \kappa_2}\right).
\label{WignerGreen}
\ee
For large $z$, only the  negative sign provides the correct asymptotic  limit $G_s(z) \sim \frac{1}{z}$. Taking the imaginary part according to  (\ref{green}), we arrive at the celebrated  Wigner semicircle distribution  of Random Matrix Theory
\be
\rho_s(\lambda)=\frac{1}{2\pi \kappa_2} \sqrt{ 4\kappa_2-\lambda^2}.
\ee
The Wigner distribution is therefore the FRV counterpart of the Gaussian distribution in the classical probability theory. 

The crucial notion in free probability  is the freeness condition which replaces independence in the classical probability calculus. 
The most intuitive definition of freeness of $\bb{A}$ and $\bb{B}$ is based on the complete decorrelation of the corresponding  eigenvectors, i.e. their eigenbases are maximally random oriented. The pair $\bb{A}$ and $\bb{UBU}^{\dagger}$ becomes free in the limit when the size of matrices tends to infinity, provided  that $\bb{U}$ is Haar unitary (i.e. infinitely large random unitary matrix from Circular Unitary Ensemble).  For a free pair $(\bb{A},\bb{B})$ a powerful addition law holds 
\be
R_{\bb{A}+\bb{B}}(z)=R_{\bb{A}}(z)+R_{\bb{B}}(z),
\label{hermadd}
\ee
linearizing the spectral problem of convolution of two non-commuting free random ensembles. 
Since the R-transforms are the generating functions for the free cumulants, the above law imposes the additivity of free cumulants, in an analogy to the addition law for the logarithms  of the characteristic functions in the classical probability theory.

It is suitable here to demonstrate the difference between the classical and FRV calculus on the simple example of convolution of two distributions, each of them composed  of a binary set of -1/2 and 1/2.  In the classical case, the resulting distribution is discrete,  where the values -1 and  1 are obtained with the probability 1/4 each, and the value 0 is obtained with the probability 1/2. The matrix-valued  version of this example could be realized in terms of two identical large matrices $\bb{A}$ and $\bb{B}$, each of them having on the diagonal the equal number of -1/2 and 1/2.  The spectral density functions are therefore given by 
\be
\rho_\bb{A}(\lambda)=\rho_\bb{B}(\lambda)=\frac{1}{2}\delta\left(\lambda+\frac{1}{2}\right)+\frac{1}{2}\delta\left(\lambda- \frac{1}{2}\right)
\label{projectors}
\ee  
or, equivalently, the Green's functions read
\be
G_\bb{A}(z)=G_\bb{B}(z)=\frac{1}{2} \left(  \frac{1}{z+\frac{1}{2}} +\frac{1}{z-\frac{1}{2}} \right).
\ee
Substituting $ z \rightarrow R(G_{\bb{A}}(z)) + 1/G_{\bb{A}}(z)$ and using the definition (\ref{Rtransform}), one arrives first at $R_{\bb{A}}(z)= \frac{\sqrt{1+z^2}-1}{2z}=R_{\bb{B}}(z)$.  Then, making use of the addition law,  $R_{\bb{A}+\bb{B}}(z)=R_{\bb{A}}(z)+R_{\bb{B}}(z)$ and using \eqref{Rtransform} but in the opposite direction, one finally arrives at 
\be
G_{\bb{A}+\bb{B}}(z)=\frac{1}{\sqrt{z^2-1}},
\label{Greenarcsin}
\ee
which, via the imaginary part, leads to the {\it continuous} spectral arsine law
\be
\rho_{\bb{A}+\bb{B}}(\lambda)=\frac{1}{2 \pi \sqrt{(1-\lambda)(1+\lambda)}}.
\label{arcsine}
\ee
One can easily check this numerically by adding large matrices $\bb{A}$ and $\bb{UBU}^{\dagger}$, where $\bb{U}$ comes from the Haar measure.  Diagonalizing the sum and plotting the histogram of the resulting eigenvalues, one obtains the spectral distribution \eqref{arcsine}. 

Surprisingly, a similar construction exists for the product of ensembles $\bb{A}$ and $\bb{B}$, despite in general the product of Hermitian (symmetric)  operators is not necessarily Hermitian.  If one of the matrices is positive (e.g. $\bb{A}$), the moments $\Tr(\bb{AB})^k$ can be rewritten under trace as $\Tr(\sqrt{\bb{A}}\bb{B}\sqrt{\bb{A}})^k$, while $\sqrt{\bb{A}}\bb{B}\sqrt{\bb{A}}$ is Hermitian by construction and shares the same eigenvalues as $\bb{AB}$. This observation leads to  the  corresponding multiplication law~\cite{BJNPRODUCTS}
\be
R_{\bb{AB}}(G_{\bb{AB}})=R_\bb{A}(G_\bb{B})R_\bb{B}(G_\bb{A}),
\label{hermmult}
\ee
where 
\be
G_{\bb{A}}=G_{\bb{AB}}R_\bb{A}(G_{\bb{B}}), \,\,\,\,\,\,\,\,\,\,\,\,\,\,  G_\bb{B}=G_{\bb{AB}}R_\bb{B}(G_\bb{A}).
\label{hermmult1}
\ee
The original construction of the multiplication law by  Voiculescu~\cite{VOICULESCU} (S-transform) is related to this formulation via  
\be
R(z)=\frac{1}{S(zR(z))}, \,\,\,\,\,\,\,\,\,\,\,\,\,\,\,\,\, S(z)=\frac{1}{R(zS(z))}, \label{Stransform}
\ee
so that $S_{\bb{AB}}(z)=S_\bb{A}(z)S_\bb{B}(z)$. Note that the formulation in the language of the S-transform  requires $R(0) \neq 0$ for both ensembles $\bb{A}$ and $\bb{B}$, which may be weakened~\cite{SPEICHERRAO}. Equations (\ref{hermmult},\ref{hermmult1}) are still valid, even if $R(0)=0$.

Finally, let us consider the important case when the resulting non-Hermitian matrix $\bb{X}$ can be decomposed as $\bb{X}=\bb{PU}$, where $\bb{P}$ is positive, $\bb{U}$ is Haar unitary and $\bb{P}$ and $\bb{U}$ are free. In such a case, the complex spectrum possesses a polar symmetry, so the spectral problem is quasi-one dimensional, i.e. only the radial spectral density $\rho(\lambda,\bar{\lambda})=\rho(|\lambda| = s)$ is non-trivial. In such circumstances the  Haagerup-Larsen theorem provides a remarkably simple relation between the radial cumulative distribution function $F(s)=2\pi \int_0^s  s' \rho(s') ds'$ and the S-transform of $\bb{P}^2$~\cite{HAAGERUPLARSEN}
\be
S_{\bb{P}^2}(F(s)-1)=\frac{1}{s^2}. \label{eq:HaagLars}
\ee
The spectrum is always confined to the ring with radii $s_{min}$,  $s_{max}$, where 
$s_{min}^{-2}=\int_{0}^{\infty}x^{-2} \rho_{\bb{P}}(x)dx$ and $s_{max}^2=\int_{0}^{\infty}x^2\rho_{\bb{P}}(x)dx$. Note that this case includes  in particular $s_{min}=0$ and $s_{max}=\infty$. This is the so-called single ring theorem~\cite{FEINBERGZEE}. Recently, it was proven~\cite{HAAGLARSVECTORS}, that the same radial cumulative distribution function $F(s)$  provides the information about certain eigenvector correlator
\be
O(s) \equiv \lim_{N \rightarrow \infty} \frac{1}{N^2} \left< \sum_{\alpha} O_{\alpha \alpha} \delta^{(2)} (\lambda-\lambda_{\alpha})\right>=\frac{F(s)(1-F(s))}{\pi s^2}, \label{eq:HaagLarsCorr}
\ee
where  $O_{\alpha \beta} = \<L_{\alpha} |L_{\beta}\>\<R_{\beta}|R_{\alpha}\>$, and $\ket{R_{\alpha}} $ and $\bra{L_{\alpha}}$ are right and left eigenvectors of the non-Hermitian matrix $\bb{X}$, respectively (see also Sec.~ \ref{sec:NonHermitianRandomMatrices}).

\subsection{Free Random Variable cookbook}
Here we present brief recipes for the addition and multiplication of large random matrices and introduce some additional FRV transforms related to moments and cumulants. They found an application in description of spectral properties of large covariance matrices in particular in the financial context~\cite{SNARSKA,SNARSKA2}. Recently, they also turned out to be crucial in the cleaning of noisy covariance matrices~\cite{BOUCHAUDREVIEW,BOUCHAUDRIE} and studying the properties of the eigenvectors of such matrices~\cite{BOUCHVECT1}.

The procedure for addition goes as follows:
\begin{enumerate}
\item Knowing individual spectral functions $\rho_\bb{A}(\lambda)$, $\rho_\bb{B}(\lambda)$, we calculate the corresponding Green's functions $G_\bb{A}(z)$ and $G_\bb{B}(z)$.
\item Using the relation (\ref{Rtransform}), we  construct $R_\bb{A}(z)$ and $R_\bb{B}(z)$,  which we then add ({\it cf.}  (\ref{hermadd})) forming $R_{\bb{A}+\bb{B}}(z)$.
\item Finally, using again the relation (\ref{Rtransform}), we reconstruct $G_{\bb{A}+\bb{B}}(z)$, which via the imaginary part yields $\rho_{\bb{A}+\bb{B}}(\lambda)$. 
\end{enumerate}
Sometimes it is practical  to use the functional inverse of the Green's function (nicknamed as 'Blue's function'), defined as 
$B[G(z)]=G[B(z)]=z$ and related to the R-transform via  $B(z)=R(z)+1/z$.   The procedure for addition is identical, except the obvious shift in the addition law  $(B_{\bb{A}+\bb{B}}(z)= B_\bb{A}(z)+B_\bb{B}(z) -1/z)$. 

The multiplication  algorithm is as follows:
\begin{enumerate}
\item Knowing individual spectral functions $\rho_\bb{A}(\lambda)$, $\rho_\bb{B}(\lambda)$, we write down the corresponding Green's functions $G_\bb{A}(z)$ and $G_\bb{B}(z)$, and then,  using the relation (\ref{Rtransform}), we  construct $R_\bb{A}(z)$ and $R_\bb{B}(z)$.   
\item Using (\ref{Stransform}), we calculate $S_\bb{A}(z)$ and $S_\bb{B}(z)$, which we then multiply (if  the aforementioned positivity condition holds), getting 
$S_{\bb{A} \cdot \bb{B}}(z)$.
\item Reversing the order of above operations, we finally reconstruct $G_{\bb{A} \cdot \bb{B}}(z)$, the imaginary part of which yields $\rho_{\bb{A} \cdot \bb{B}}(\lambda)$. 
\end{enumerate}
It turns out that in some instances it is more practical to use the moment generating function  $M(z) \equiv z G(z)-1$ and its functional inverse, defined as $M[N(z)]=N[M(z)]=z$.  The procedure for the multiplication reads then $N_\bb{A}(z) N_\bb{B}(z)= \frac{1+z}{z} N_{\bb{A} \cdot \bb{B}} (z)$, since $S(z)=\frac{1+z}{z} \frac{1}{N(z)}$. 

We list in  the form of the Table~\ref{tab:WishartAntiwishart} the corresponding transforms for the Wishart  $\bb{W}= \frac{1}{T} \bb{xx}^{\dagger} $ (represented by $N \times N$ matrix)  and  the anti-Wishart matrix $\bb{aW}\equiv \frac{1}{N} \bb{x}^{\dagger}\bb{x}$ (represented by $T \times T$ matrix). Note the duality relation $M_\bb{W}(z)=rM_{\bb{aW}}(rz)$.

\begin{table}[h]
\begin{center}
 \begin{tabular}{||c c c ||} 
 \hline
  & Wishart  & anti-Wishart   \\ 
  \hline\hline
  
 Matrix form  & $\frac{1}{T}\bb{xx}^{\dagger}$ & $\frac{1}{N} \bb{x}^{\dagger}\bb{x} $
  \\  
 \hline
 
 R-transform &  $R_\bb{W}(z)=\frac{1}{1-rz}$ & $R_{\bb{aW}}(z)= \frac{r}{r-z}$ 
  \\ 
 \hline
 
 S-transform & $S_\bb{W}(z) = \frac{1}{1+rz}$ & $S_{\bb{aW}}(z)=\frac{r}{r+z}  $ 
 \\ 
 \hline
 
N-transform   & $N_\bb{W}(z)=\frac{1+z}{z} (1+rz) $\,\,\,\,\,\,\,\,   & $N_{\bb{aW}}(z)=  \frac{(1+z)(r+z)}{rz}$ 
\\ 
 \hline
 \hline 
\end{tabular}
\caption{Comparison between Wishart and anti-Wishart transforms. \label{tab:WishartAntiwishart}}
\end{center}
\end{table}

\section{Unit time-lagged correlation matrices --  quasi-one dimensional reductions}  \label{sec:FRV2}

We consider the Pearson estimator of the autocorrelation matrix with the time shift (lag) of $\tau$ units
\be
C_{ij}^{\tau}=\frac{1}{T-\tau}\sum_{t=1}^{T-\tau} x_{it} \bar{x}_{j,t+\tau},
\ee
where $x_{it}$ represents measurements of $i=1,...,N$ objects at times $t=1,...,T$. First, we are interested in the unit time lag ($\tau=1$) in the limit when both $N$ and $T$ tend to infinity, keeping their ratio $r=N/T$ fixed (we choose $r \leq 1$).  Note, that in matrix notation 
\be
 \bb{C}=\frac{1}{T} \bb{xDx}^{\dagger},
\ee
with $D_{tt'}=\delta_{t+\tau,t'}$. For $T\to\infty$ and fixed $\tau$ we can put $T\sim (T-1)$. Moreover, in this section we associate $T+k$ with $k$ in the argument of the Kronecker delta in order to secure the existence of the inverse matrix for $\bb{D}$. We call this cyclic approximation, because $\bb{D}$ is then a representation of a cyclic permutation. The accuracy of such an approximation has to be checked numerically a posteriori.

Higher values of $\tau$, which we also refer to the depth of the lag, are considered later in Section \ref{sec:Analysis}, devoted to non-Hermitian matrices, where the non-trivial limiting procedure $N,T,\tau \to \infty$ with $N/T$ and $\tau/T$ fixed is studied.

\subsection{Symmetrization}
Symmetrization of the lag matrix $\bb{D}$ leads to $\bb{D}^{sym}=\frac{1}{2}(\bb{D}+\bb{D}^T)$.  The eigenvalues of the symmetrized matrix read $\lambda_k=\cos (\pi k /(T+1))$ and, in the limit $T \rightarrow \infty$, $\tau$ fixed, the resolvent can be approximated as 
\be
G_{\bb{D}^{sym}}(z)=\int_0^1 dx \frac{1}{z- \cos (\pi x)}=\frac{1}{\sqrt{z^2-1}},
\ee
 which is identical to the  resolvent for the free arcsine law \eqref{Greenarcsin}. Inserting the above result into the chain rule for the S-transforms~\cite{USREVIEW},  one obtains  the fourth order polynomial equation (Ferrari equation) for the moment generating function $M(z)$
\be
r^3 M^4 +2r^2(1+r)M^3 +r(1+4r+r^2-z^2)M^2 +2(r^2+r-z^2)M  +r=0. \label{eq:quarticM}
\ee
This equation was first presented without a proof in \cite{HINDU}  and derived in~\cite{USREVIEW}.

\subsection{Spectral whitening}
We represent the lagged correlation matrix  as $\bb{C}^{\tau}=\frac{1}{T} \bb{xy}^{\dagger}$, where $y_{it}=x_{i,t+\tau}$. Then we calculate its 'squared modulus'  
 $\bb{CC}^{\dagger}=T^{-2} \bb{xy}^{\dagger}\bb{yx}^{\dagger}$. The spectral content of such a matrix  is complicated   - the real eigenvalues not only capture the cross-correlations between $x$ and $y$, but also reflect the inner-correlation within the 'Wishart-like' ensembles  $\bb{W_x}=\frac{1}{T}\bb{xx}^{\dagger}$ and  $\bb{W_y}=\frac{1}{T} \bb{yy}^{\dagger}$. In order to disentangle these two sources of information, we remove by hand  all the correlations  within these two Wishart ensembles. We diagonalize the equal-time correlation matrix $\bb{W_x}=\bb{U \Lambda U}^{\dagger}$ and define a new time series $\bb{x}'=\bb{\Lambda}^{-1/2}\bb{U}^{\dagger}\bb{x}$,  which has the property that its equal-time correlation matrix is the identity.  This procedure, applied likewise to $\bb{y}$, is known as whitening.  The next step uses the observation that the moments of the square involve the product of two whitened anti-Wishart type ensembles $\bb{W}_1=\frac{1}{T}\bb{x}'^{\dagger}\bb{x'}$ and $\bb{W}_2=\frac{1}{T} \bb{y}'^{\dagger}\bb{y}'$.  Since both ensembles are now represented by $T \times T$ matrices, and the nonzero-eigenvalues of both Wishart and anti-Wishart ensembles are identical and equal to 1,  the remaining $T-N$ eigenvalues have to be zeroes. 
 In the absence of the time-lagged correlations, the spectral problem reduces therefore to the free multiplication of two projectors, the Green's functions of which read
 $G_{\bb{W}_1}(z)=G_{\bb{W}_2}(z)=r \frac{1}{z-1}+(1-r)\frac{1}{z}$. A brief calculation (see Appendix \ref{App:FreeProjectors}) yields the quadratic equation for M
 \be
 (z-1)M^2+(z-2r)M-r^2=0.
 \ee
 The resulting spectral density  belongs to the family of free Jacobi measures~\cite{VOICULESCU}. This method for  looking for general cross-correlations in economic data was proposed in~\cite{BOUCHAUD}.  
 
 \subsection{Abelization}

A way to deal with the non-Hermitian matrices with circularly symmetric spectrum was proposed by Biely and Thurner \cite{THURNERBIELY} with no proof and without specifying the class of matrices which this procedure applies to. They argued that the marginal distribution $\rho_x(x)=\int\rho(z=x+iy)dy$ of such a matrix is related to the spectral density of a symmetrized problem $\bb{C}^{{\rm sym}}=\frac{1}{2}(\bb{C}^{\tau}+(\bb{C}^{\tau})^{\dagger})$ via $\rho_x(x)=\rho^{sym}(\sqrt{2}x)$. Later, it was shown \cite{BURDAJANIKWACLAW} that this property does not hold for the product of two independent GUE matrices.

We argue here that the Abelization  property holds for normal matrices $\bb{X}$, i.e. those which can be diagonalized by a unitary transformation $\bb{X}=\bb{U\Lambda U}^{\dagger}$. The Hermitian conjugate $\bb{X}^{\dagger}$ can be diagonalized by the same matrix $\bb{U}$, thus the eigenvalues of the symmetrized matrix $\bb{C}^{{\rm sym}}=\frac{1}{2}\bb{U}(\bb{\Lambda}+\bb{\Lambda}^{\dagger})\bb{U}^{\dagger}$ are the real parts of the eigenvalues of $\bb{X}$, thus $\rho_{x}(x)=\rho^{{\rm sym}}(x)$.

If the spectrum is circularly symmetric, $\rho(z,\zb )=\rho(s)$, where $s=|z|$, the marginal distribution can be rewritten as the Abel transform of the spectral density:
\begin{equation}
\rho_x(x)=2\int\limits_{x}^{\infty}\frac{s \rho(s) ds}{\sqrt{s^2-x^2}}.
\end{equation}
Knowing the spectral density of the symmetrized problem, one can recover the eigenvalue spectrum of the initial matrix via the inverse Abel transform
\begin{equation}
\rho(s)=-\frac{1}{\pi}\int\limits_{s}^{\infty}\frac{d\rho^{sym}(x)}{dx}\frac{dx}{\sqrt{x^2-s^2}}.
\end{equation}
The spectral density of the symmetrized problem is obtained from \eqref{eq:quarticM}.

Unfortunately, the lagged correlation matrices are usually not normal, so a different approach has to be used to get the proper form of the radial spectral density.


\begin{figure}
\includegraphics[width=\textwidth]{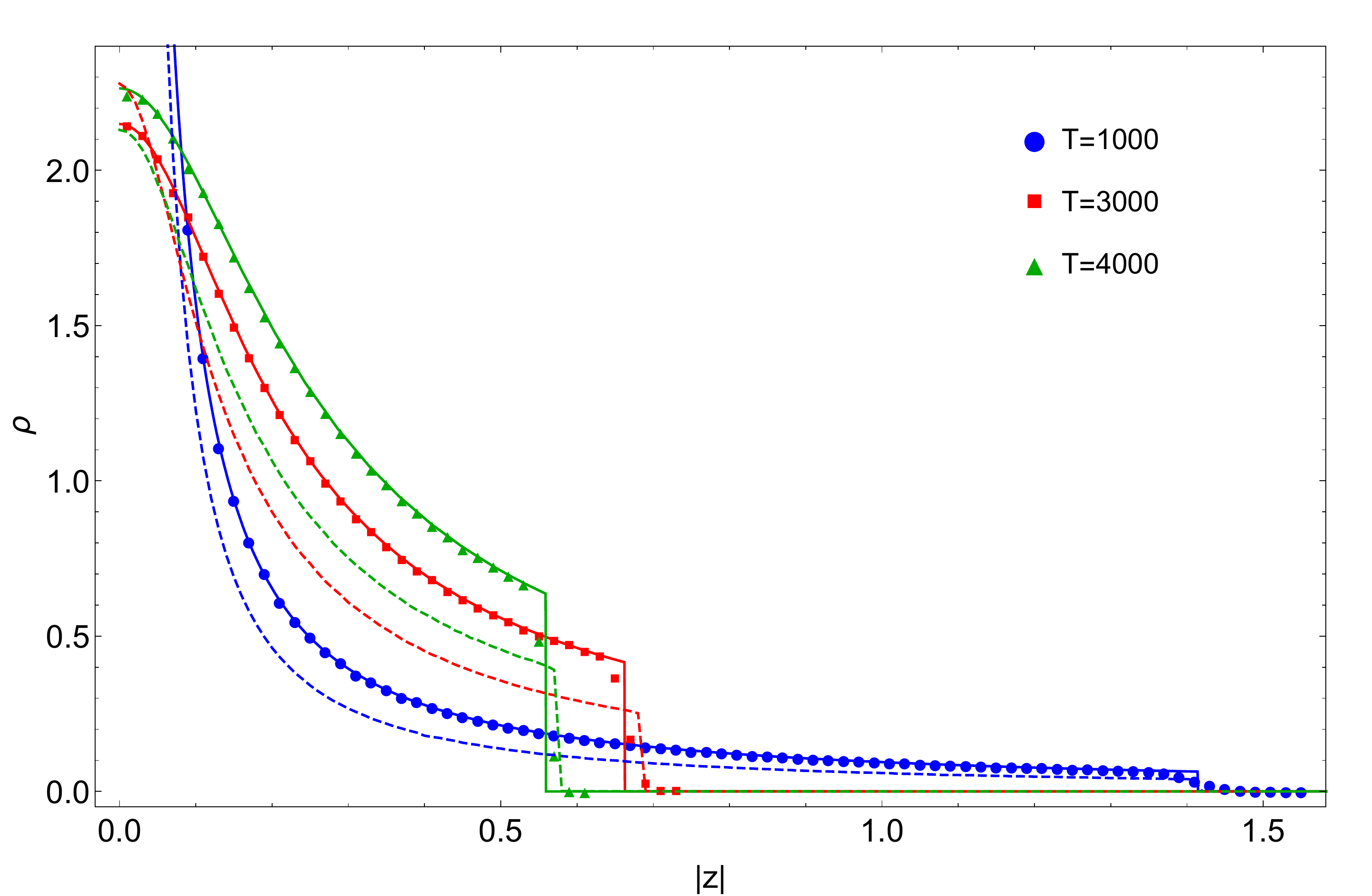}
\caption{A numerical diagonalization of 1000 matrices of size $N=1000$ and different rectangularities. The solid lines present the solution obtained from the Haagerup-Larsen theorem, while the dashed line from the Abelization, where the only known analytical result corresponds to $r=1$. The agreement of the numerical results with the solution obtained from the Haagerup-Larsen theorem shows the validity of the cyclic approximation. The discrepancies at the edges are the effects of finite size of matrices.
\label{numerics1}}
\end{figure}

\subsection{Exact radial spectrum from the Haagerup-Larsen  theorem }  \label{sec:HaagLar}

We work in the cyclic approximation and $\bb{D}$ is a particular case of the circulant (permutation) matrix, its $T$ eigenvalues are just all complex roots of unity, i.e.  $\lambda_t =\exp(2\pi i \frac{t}{T})$. For large $T$ the eigenvalues tend to the uniform distribution on the unit circle, i.e. they approach the  constant measure  (Haar measure) of Circular Unitary Ensemble (CUE). Due to the Sylvester's determinant identity the matrices $\frac{1}{T}\bb{xDx}^{\dagger}$ and $\frac{1}{T}\bb{x}^{\dagger}\bb{xD}$ share the same non-zero eigenvalues, therefore the spectral problem of the time-lagged correlation matrix $\bb{C}$ is equivalent to the product of CUE ensemble and the anti-Wishart-type ensemble $\frac{1}{T} \bb{x}^{\dagger}\bb{x}= r \frac{1}{N} \bb{x}^{\dagger}\bb{x}$. The probability density function of the anti-Wishart ensemble is invariant under the unitary transformations, therefore its eigenbasis is already randomly oriented. The two matrices in the product are mutually free and we are allowed to use the Haagerup-Larsen theorem.

Since under the rescaling of an arbitrary matrix $\bb{X}$ by a real number $r$, $R_{r\bb{X}}(z)=rR_\bb{X}(rz)$, the R transform for  the matrix  $\bb{Y}={\frac{1}{T} \bb{x}^{\dagger}\bb{x}}$ reads $R_\bb{Y}(z)=rR_{\bb{aW}}(rz)= r \frac{r}{r-(zr)}= \frac{r}{1-z}$. We calculate its Green's function using \eqref{Rtransform}, which takes the form
\begin{equation}
G_{\bb{Y}}(z)=\frac{1-r+z+\sqrt{(1-r+z)^2-4z}}{2z}.
\end{equation}
The resolvent of a square of a matrix can be expressed by the Green's function of the matrix itself, with the help of the identity
\begin{equation}
G_{\bb{Y}^2}(z^2)=\frac{1}{2z}G_{\bb{Y}}(z)-\frac{1}{2z}G_{\bb{Y}}(-z),
\end{equation}
 which follows form the partial fraction decomposition. The S-transform, calculated via \eqref{Stransform} reads
 \begin{equation}
 S_{\bb{Y}^2}(z)=\frac{r+1+z}{(r+z)(r+2z+1)^2}.
 \end{equation}
 Using the Haagerup-Larsen theorem \eqref{eq:HaagLars}, we finally arrive at the cubic equation for the radial cumulative distribution function
 \begin{equation}
 4F^3+8F^2(r-1)+F(5(r-1)^2-s^2)+(r-1)^3-rs^2=0. \label{CDFequation}
 \end{equation} 
 
 Upon choosing a real valued branch of the solution and taking into account the absence of zero modes in the original problem, one can calculate the spectral density
 \begin{equation}
 \rho(s)=\frac{1}{2\pi s r}\frac{dF}{ds}.
 \end{equation}
 As a shortcut, we also deduce from (\ref{CDFequation}) the values of the spectral radii (corresponding to $F=1$ and $F=0$)
\begin{equation}
s_{ext}=\sqrt{r(r+1)}, \qquad s_{int}=\frac{(r-1)^{3/2}}{\sqrt{r}}\theta(r-1), \label{eq:spectralradii}
\end{equation} 
 where $\theta(x)$ is the Heaviside theta function. We remark that the same result can be obtained via calculation of the second and inverse second moments of the spectral density of anti-Wishart, as stated in the Haagerup-Larsen theorem. Several solutions compared to the numerical simulations are plotted in Fig. \ref{numerics1}. This result confirms the  diagrammatic calculation by~\cite{JAROSZLAG}.

\section{True-lagged correlation matrices and eigenvector correlators} \label{sec:diagrammatic}
\subsection{Non-hermitian random matrices}\label{sec:NonHermitianRandomMatrices}

All the above-mentioned approaches have  the same feature - they reduce the complex spectral problem to the calculation of real eigenvalues, either by imposing real spectra (symmetrization, Abelization) or by reducing the two-dimensional complex spectrum to a quasi-one dimensional case, imposing  azimuthal symmetry and considering only the radial variables (whitening, Haagerup-Larsen theorem). Moreover, securing the existence of the inverse of the delay matrix by adding additional non-zero elements (cyclic approximation) does not allow to probe the spectrum with varying lag depth, because the eigenvalues of such $\bb{D}$ remain the same, regardless of $\tau$.

In order to attack the truly non-Hermitian problem, we first have to find a representation of the complex Dirac delta, since the Sochocki-Plemelj formula ceases to work in the complex case. In the spirit of the electrostatic analogy, the applications of which to Random Matrix Theory were very fruitful, we use the Poisson kernel in two dimensions
\begin{equation}
\delta^{(2)}(z)=\lim_{\epsilon\to 0}\frac{1}{\pi}\frac{\epsilon^2}{(|z|^2+\epsilon^{2})^{2}}=\lim_{\epsilon\to 0} \frac{1}{\pi}\partial_{\zb} \frac{\zb}{|z|^2+\epsilon^2}.
\end{equation}
The differentiation with respect to $\zb$ reduces the power of the denominator, which simplifies the object which one has to deal with. Let $\bb{X} $ be a random matrix, the mean spectral density of which we want to calculate. Denoting
\begin{equation}
g(z,\zb,w,\wb)=\< \frac{1}{N}\Tr \frac{\zb \idm_{N} -\bb{X}^{\dagger}}{(z \idm_N-\bb{X})(\zb \idm_{N}-\bb{X}^{\dagger})+|w|^2 \idm_{N}}\>, \label{eq:gDef}
\end{equation}
the distribution of the eigenvalues on the complex plane can be calculated from~\cite{SOMMERS,FYODOROVSOMMERS,BROWN}
\begin{equation}
\rho(z,\zb)=\frac{1}{\pi}\lim_{|w|\to 0} \partial_{\zb} g(z,\zb,w,\wb).
\end{equation}

However, due to the nonlinearity in the denominator, it is very challenging to calculate $g$. It was realized in \cite{JANIKNOWAK,JARNOW}, that $g$ can be considered as a part of an extended object which is linear in $\bb{X}$ and therefore is easier accessible in practical calculations. Consider a $2N\times 2N$ matrix
\begin{equation}
\cG=\<\left(\begin{array}{cc}
z\idm_{N} -\bb{X} & i\wb \idm_{N} \\
iw \idm_{N} & \zb \idm_{N} -\bb{X}^{\dagger}
\end{array}\right)^{-1}\> = \< \left[ Q\otimes \idm_{N} - \cX\right]^{-1}\>,
\end{equation}
where $Q$ is a $2\times 2$ matrix representation of a quaternion and $\cX$ is a duplicated matrix
\begin{equation}
Q=\left(\begin{array}{cc}
z & i\wb \\ 
iw & \zb 
\end{array}\right), \qquad 
\cX=\left(\begin{array}{cc}
\bb{X} & 0 \\
0 & \bb{X}^{\dagger}
\end{array}\right). \label{eq:BlockStructure}
\end{equation}
We remark that in the literature on non-Hermitian random matrices there exists another representation of a quaternion where the off-diagonal elements are devoid of a factor of $i$ and the element above the diagonal appears with a minus sign. These representations are isomorphic, however throughout this paper we adapt the convention with imaginary units, which makes equations more symmetric. 

The block structure reflects a compound nature of the considered object $\cG$. We therefore adopt a convention of writing its elements which is compatible with this structure and in terms of indices we write $\cG^{ij}_{\alpha\beta}$. The subscript Greek indices refer to the quaternionic nature, run from $1$ to $2$ and enumerate blocks of the duplicated matrix. Latin indices in the superscript enumerate elements of the matrix within each block. By taking the partial trace over the space of matrix indices (which is a block trace of the duplicated matrix) we produce a $2\times 2$ matrix which is itself a quaternion and we refer to this as the quaternionic Green's function (generalized Green's function)
\begin{equation}
G(Q)=\frac{1}{N}\btr \cG=\left(\begin{array}{cc}
\<\frac{1}{N}\Tr(\zb\idm_N-\bb{X}^{\dagger})\bb{D}^{-1}\> & \<\frac{-i \wb}{N} \Tr \bb{D}^{-1} \>\\
\<\frac{-i w}{N} \Tr \bb{D}^{-1} \> & \<\frac{1}{N}\Tr (z\idm_N-\bb{X})\bb{D}^{-1} \>
\end{array}\right),
\end{equation}
where  $\bb{D}=(z\idm_N-\bb{X})(\zb\idm_N-\bb{X}^{\dagger})+|w|^2\idm_{N}$. In the analogy to the complex resolvent for Hermitian matrices, we give $\cG$ the name quaternionic resolvent. We identify the upper-left element of $G$ with \eqref{eq:gDef} and denote all its elements
\begin{equation}
G(Q)=\left(\begin{array}{cc}
g & i\vb \\
i v & \gb 
\end{array}\right).
\end{equation} 
The lower-diagonal element is just a complex conjugate copy of $g$, however, the off-diagonal elements in the $N\to \infty$ limit endow us for free with additional information.

If a non-Hermitian matrix is diagonalizable, it possesses a set of left $\bra{L_i}$ and right $\ket{R_i}$ eigenvectors, solving the eigenproblem
\begin{equation}
\bb{X}\ket{R_i}=\lambda_i \ket{R_i}, \quad \bra{L_i}\bb{X}=\bra{L_i} \lambda_i.
\end{equation}
They form a  biorthogonal set, i.e. they are normalized by the condition $\braket{L_i}{R_j}=\delta_{ij}$, however the left and right eigenvectors are not orthogonal among themselves, $\braket{L_i}{L_j}\neq \delta_{ij}\neq\braket{R_i}{R_j}$. The biorthogonality condition leaves a freedom of rescaling the eigenvectors by an arbitrary complex number $\ket{R_i}\to c_{i}\ket{R_i}$, $\bra{L_i}\to \bra{L_i}c_i^{-1}$, and also of multiplying by a unitary matrix $\ket{R_i}\to \bb{U}\ket{R_i}$, $\bra{L_i}\to\bra{L_i}\bb{U}^{\dagger}$. The simplest non-trivial quantity which is invariant under these transformations is the matrix of overlaps  $O_{\alpha\beta}=\braket{L_\alpha}{L_\beta}\braket{R_{\alpha}}{R_{\beta}}$, introduced by Chalker and Mehlig~\cite{CHALKERMEHLIG,CHALKERMEHLIG2}. The diagonal elements are the squares of the eigenvalue condition numbers, which play a significant role in the stability of the spectrum of non-normal matrices~\cite{WILKINSON,TREFETHEN,BERRY}. Chalker and Mehlig introduced a one point correlation function associated with the diagonal part of the overlap matrix
\begin{equation}
O_N(z)=\<\frac{1}{N^2} \sum_{i=1}^{N}O_{ii}\delta^{(2)}(z-\lambda_i)\>. \label{eq:Correlator}
\end{equation}
It naturally appears in non-Hermitian systems such as open chaotic scattering~\cite{SAVINSOK,FYODSAV,FYODOROVMEHLIG,BEENAKKER,BEENAKKER2} and also in diffusion on large non-Hermitian matrices~\cite{DIFFPRL,DIFFNPB}.

In the large $N$ limit the product of the off-diagonal elements of the quaternionic Green's function reproduces the eigenvector correlator \cite{HAAGLARSVECTORS,JANIKNOWAKCORR}
\begin{equation}
O(z)\equiv \lim_{N\to\infty} O_N(z)=\lim_{N\to\infty}\lim_{|w|\to 0} \frac{1}{\pi} |v|^2.
\end{equation}

A huge benefit of the linearization formalism comes from the fact that $\cG$ can be expanded into a geometric series in $\cQ=Q\otimes \idm_{N}$ and $\cX$
\begin{equation}
\cG=\cQ^{-1}+\<\cQ^{-1}\cX\cQ^{-1}\>+\<\cQ^{-1}\cX\cQ^{-1}\cX\cQ^{-1}\>+\<\cQ^{-1}\cX\cQ^{-1}\cX\cQ^{-1}\cX\cQ^{-1}\>+\ldots, \label{eq:MomentExpansion}
\end{equation}
provided that $||\cX\cQ^{-1}|| < ||\idm_{2N}||$ in some norm. If we consider $z$ lying outside the spectrum of $\bb{X}$, then we are allowed to set $|w|$ to 0 and, as a consequence, $g$ is a holomorphic function. However, inside the spectrum we need to keep $w$ sufficiently large so that the expansion \eqref{eq:MomentExpansion} is legible. At the end of the calculations we set $|w|\to 0$ and then $g$ will depend both on $z$ and $\zb$, indicating the non-vanishing spectral density.

The quaternionic Green's function, obtained as a block trace of \eqref{eq:MomentExpansion}, generates all mixed moments of non-Hermitian matrices in a similar manner as the complex Green's function for Hermitian matrices \eqref{eq:HermMoments}. Now, due to the fact that in general $\bb{X}$ does not commute with $\bb{X}^{\dagger}$, the class of possible moments is much broader. To show how they are encoded, we consider a quaternionic moment generating function, which we define as $M(Q)=Q^{-1}G(Q^{-1})Q^{-1}-Q^{-1}$, and its expansion into moments of $Q$
\begin{equation}
M(Q)=\<\frac{1}{N}\btr \cX\>+\<\frac{1}{N}\btr \cX \cQ \cX\>+\<\frac{1}{N}\btr \cX\cQ\cX\cQ\cX \>+\ldots \,\, . \label{eq:MMomentExP}
\end{equation}
$M$ is itself a $2\times 2$ matrix, the entries of which together with the entries of the quaternion we  denote here in a unifying way
\begin{equation}
M(Q)=\left(\begin{array}{cc}
M_{11} & M_{1\bar{1}} \\
M_{\bar{1}1} & M_{\bar{1}\bar{1}}
\end{array}\right),\quad 
Q=\left(\begin{array}{cc}
Q_{11} & Q_{1\bar{1}} \\
Q_{\bar{1}1} & Q_{\bar{1}\bar{1}}
\end{array}\right).
\end{equation}

Suppose now that we would like to extract from $M(Q)$ the mixed moment $\<\frac{1}{N}\Tr \bb{X}^{a_1}\bb{X}^{a_2}\ldots\bb{X}^{a_n}\>$, where each each of $a$'s is either $1$ or $\dagger$. The mnemotechnical rule, originating from the expansion \eqref{eq:MMomentExP}, goes as follows. First, one has to associate $\dagger$ with $\bar{1}$ and, obviously, $1$ with $1$. Then, identify first and last terms in the chain and consider the component $a_1a_n$  of $M$. In the next step one expands the appropriate component in powers of $z,\zb,iw,i\wb$ and reads out the coefficient of $Q_{a_1a_2}Q_{a_2a_3}\ldots Q_{a_{n-1}a_n}$. 

As an example we consider $\<\frac{1}{N}\Tr \bb{XX}^{\dagger}\bb{X}^{\dagger}\bb{X}\>$. We have to expand $M_{11}$ into a series and consider the coefficient of $Q_{1\bar{1}}Q_{\bar{1}\bar{1}}Q_{\bar{1}1}=(i\wb) \zb (iw)$. The matrix $Q$ plays the role of a transition matrix in this chain and one can easily convince himself that all possible mixed moments are encoded in this way. We also remark that the quaternionic R-transform generates all possible mixed cumulants (a.k.a. connected moments) in the same way as $M$ generates all mixed moments~\cite{BURDASWIECH}. A particular type of mixed moments and its relation to the eigenvector correlation function $O(z)$ was studies thoroughly in~\cite{WALTERSSTARR}.

\subsection{Moment expansion of the quaternionic Green's function}

In this subsection we shall take the problem of determining the quaternionic Green's function of the lagged correlation matrix by expanding it into a power series and calculating the expectations of each terms. Then we find a way to sum all terms and write the Green's function in a neat form. 

We remind that the lagged correlation matrix can be written as a matrix multiplication $\bb{C}^{\tau}=\frac{1}{T-\tau} \bb{X} \bb{D}\bb{X}^{\dagger}$. To find the spectrum of $\bb{C}^{\tau}$ and analyze its properties as the depth of the lag varies, we solve a more general problem. We consider a matrix $\bb{Y}=\frac{1}{T}\bb{XAX}^{\dagger}$, where $\bb{A}$ is an arbitrary $T\times T$ matrix independent of $\bb{X}$, with no symmetry constraints and not necessary invertible. The elements of $\bb{X}$ are complexes with both real and imaginary parts Gaussian random variables of $0$ mean and $1/2$ variance. The probability density function of $\bb{X}$ can be written in a concise form $P(\bb{X})\sim \exp(-\Tr \bb{XX}^{\dagger})$. We are interested in the limit $T,N\to\infty$ with $r=N/T$ fixed, thus our results are valid also in the case when $\bb{X}$ is a real Gaussian matrix. However, for finite $N$ there is an accumulation of order of $N^{1/2}$ eigenvalues on the real line \cite{EDELMAN}.

We shall be performing calculation on $2N\times 2N$ matrices of the block structure \eqref{eq:BlockStructure}. The main object of interest is the quaternionic resolvent
\begin{equation}
\cG=\<[\cQ-\cY]^{-1}\>,
\end{equation}
where $\cQ=Q\otimes \idm_{N}$ and 
\begin{equation}
\cY=\left(\begin{array}{cc}
\frac{1}{T}\bb{XAX}^{\dagger} & 0 \\
0 & \frac{1}{T}\bb{XA}^{\dagger}\bb{X}^{\dagger}
\end{array} \right).
\end{equation}
There emerges a surprising symmetry stemming from the sandwich structure of $\bb{Y}$ which allows us to factorize $\cY=\frac{1}{T}\cX\cA\cX^{\dagger}$, with
\begin{equation}
\cX=\left(\begin{array}{cc}
\bb{X} & 0 \\
0 & \bb{X}
\end{array}\right), \qquad
\cA=\left(\begin{array}{cc}
\bb{A} & 0 \\
0 & \bb{A}^{\dagger}
\end{array}\right), \qquad 
\cX^{\dagger}=\left(\begin{array}{cc}
\bb{X}^{\dagger} & 0 \\
0 & \bb{X}^{\dagger}
\end{array}\right).
\end{equation}

Since the real and imaginary parts of elements of $\bb{X}$ are centered Gaussian random variables, the expectation of higher moments can be calculated with the knowledge of the second cumulants only by means of Wick's (Isserlis') theorem. This property transfers simply to the language of the matrix elements and the second cumulants read
\begin{equation}
\<X_{it}X_{js}\>=0=\<X_{ti}^{\dagger}X_{sj}^{\dagger}\>, \qquad \<X_{it}X^{\dagger}_{sj}\>=\delta_{ij}\delta_{ts}.
\end{equation}
This brings an additional constraint in the moment expansion of $\cG$ that the only non-vanishing pairings are those where $\bb{X}$ is paired with $\bb{X}^{\dagger}$, thus the cumulants in the moment expansion are
\begin{equation}
\< \cX_{\alpha\beta}^{at}\cX^{\dagger sb}_{\mu \nu}\>=\delta_{\alpha\beta}\delta_{\mu \nu} \delta^{ab}\delta^{st}, \qquad \<\cX_{\alpha\beta}^{at}\cX_{\mu\nu}^{sb} \>=0 = \<\cX^{\dagger at}_{\alpha\beta}\cX_{\mu\nu}^{\dagger sb}\> . \label{eq:cumulants}
\end{equation}

The expansion of $\cG$ reads
\begin{equation}
\cG^{ab}_{\alpha\beta}=(\cQ^{-1})^{ab}_{\alpha\beta}+\<\frac{1}{T}(\cQ^{-1})^{ac}_{\alpha\gamma}\cX^{cs}_{\gamma\delta}\cA^{st}_{\delta\epsilon}\cX^{\dagger td}_{\epsilon\varphi}(\cQ^{-1})^{db}_{\varphi\beta}\>+\ldots \,\, . \label{eq:Gexpansion}
\end{equation}
For the clarity of our formulas we adopt a convention that whenever two indices are repeated, we sum over the entire range of their variability.

Having \eqref{eq:cumulants}, we reduce any expectation involving strings of $\cX$'s to the products of Kronecker deltas. Although the difficulty of calculating moments is avoided, the number of terms in the sum grows as $n!$, where $n$ is the order of expansion. Moreover, even for large $T$ some of the higher order terms give a contribution of order $\mathcal{O}(1)$, therefore one has to sum all terms carefully. In order to cope with the proliferation of indices, which in fact obscure the existing structure of the expansion \eqref{eq:Gexpansion}, we represent formulas as diagrams. The details of the diagrammatic expansion we present in Appendix \ref{sec:Diagrams}, where we thoroughly derive equations stemming from the pictorial structure. Here we briefly describe the procedure of calculations. 

The moment expansion \eqref{eq:MomentExpansion} is a particular instance of the t'Hooft $1/N$ expansion in field theories with an internal $U(N)$ (or $O(N)$) group symmetry~\cite{THOOFT}. He showed that in the large $N$ limit only planar diagrams give the non-vanishing contribution, which simplifies the structure of diagrammatic expansion significantly.

Among planar diagrams produced by expressions in \eqref{eq:Gexpansion} we distinguish  a class of one-line irreducible (1LI, in physics literature they are also known as one-particle irreducible) diagrams. We denote $\Sigma^{ab}_{\alpha\beta}$ a sum over all possible 1LI planar diagrams with amputated external legs and refer to it as self-energy. These diagrams are the building block of $\cG$, which can be expressed as a series, the $n-$th term of which contains $n$ 1LI diagrams. This relation is known as the Schwinger-Dyson equation and reads
\begin{equation}
\left(\cQ_{\alpha\gamma}^{ac}-\Sigma^{ac}_{\alpha\gamma}\right)\cG^{cb}_{\gamma\beta}=\delta^{ab}\delta_{\alpha\beta}. \label{eq:SD1}
\end{equation} 

Following the ideas from \cite{BURDAJURKIEWICZ}, we introduce a dual $2T\times 2T$ matrix, which differs from $\cY$ by a cyclic permutation of factors. It is defined as $\hat{\cY}=\frac{1}{T}\cX^{\dagger}\cX\cA$. We distinguish all matrices and functions corresponding to the dual problem by putting a hat over them. We remark that contrary to the instance in \cite{BURDAJURKIEWICZ}, the dual matrix is only an auxiliary object and it does not correspond to any non-Hermitian matrix problem because the diagonal blocks are not Hermitian conjugates of each other.

We consider a moment expansion of the dual problem
\begin{equation}
\hat{\cG}=\<\left[ Q\otimes \idm_{T}-\hat{\cY}\right]^{-1}\>,
\end{equation}
for which we can write the Schwinger-Dyson equation as in the previous case
\begin{equation}
\left(Q_{\alpha\gamma}\delta^{tr}-\hat{\Sigma}^{tr}_{\alpha\gamma}\right)\hat{\cG}^{rs}_{\gamma\beta}=\delta^{ts}\delta_{\alpha\beta}. \label{SD2}
\end{equation}

With the help of the dual problem we can relate $\cG$ and $\hat{\cG}$ with self-energies, obtaining a closed system of equations. An analysis of the corresponding diagrams yields
\begin{eqnarray}
\Sigma_{\alpha\beta}^{ab}&=&\frac{1}{T}\delta^{ab}\cA_{\alpha\gamma}^{cd}\hat{\cG}_{\gamma\mu}^{dc} Q_{\mu\beta}, \\
\hat{\Sigma}_{\alpha\beta}^{ts}&=&\frac{1}{T}Q_{\alpha\gamma}\cG^{cc}_{\gamma\mu}\cA^{ts}_{\mu\beta}.
\end{eqnarray}

Substitution of self-energies to the Schwinger-Dyson equations gives us two coupled matrix equations

\begin{eqnarray}
\left(Q_{\alpha \nu}\delta^{ac}-\frac{1}{T}\delta^{ac}\cA^{ts}_{\alpha\gamma}\hat{\cG}_{\gamma\mu}^{st}Q_{\mu\nu}\right)\cG_{\nu\beta}^{cb}=\delta_{\alpha\beta}\delta^{ab},  \label{eq:MatrixCoupledFirst} \\
\left(Q_{\alpha\nu}\delta^{tr}-\frac{1}{T}Q_{\alpha\gamma}\cG^{cc}_{\gamma\mu}\cA^{tr}_{\mu\nu}\right)\hat{\cG}^{rs}_{\nu\beta}=\delta_{\alpha\beta}\delta^{ts}.
\end{eqnarray}

Keeping in mind that $G_{\alpha\beta}=\frac{1}{N}\cG^{aa}_{\alpha\beta}$ is the quaternionic Green's function, equation \eqref{eq:MatrixCoupledFirst} imples that $\cG=G\otimes \idm_{N}$. Eliminating $\hat{\cG}$ from the equations above, we obtain the final equation for the quaternionic Green's function, written in the matrix form
\begin{equation}
\left[Q-\frac{1}{T}\btr\left(\cA[\idm_{2T}-r(G\otimes\idm_{T})\cA]^{-1}\right)\right]G=\idm_{2}. \label{eq:Solution}
\end{equation}
 This equation is exact in the limit $N,T\to\infty$ with $N/T=r$ constant, although for finite size of matrices there are $1/N$ corrections. Simple algebraic manipulations  bring this equation to the following form
\begin{equation}
[Q-M_{\bb{A}}(rG)]G=\idm_{2},
\end{equation} 
where $M_\bb{A}(Q)$ is the quaternionic moment generating function of $\bb{A}$ \eqref{eq:MMomentExP}, showing that the spectral properties of $\frac{1}{T}\bb{XAX}^{\dagger}$ are completely determined by the mixed moments of $\bb{A}$.
 If additionally the matrix $\bb{A}$ is invertible, defining $G_{\bb{A}^{-1}}=\frac{1}{T}\btr (Q\otimes \idm_T-\cA^{-1})^{-1}$, the quaternionic Green's function of $\bb{A}^{-1}$, the above equation can be written in another neat form
 \begin{equation}
 \left[Q+G_{\bb{A}^{-1}}(rG)\right]G=\idm_{2}. \label{eq:SolutionSimplified}
 \end{equation}

\section{Analysis of the spectrum of the lagged correlation matrix} \label{sec:Analysis}
\subsection{Unit time lag}

We consider a unit time shift $\tau=1$ and the standard limits $N,T\to\infty$ with $r=N/T$ fixed. As in Section \ref{sec:HaagLar} we replace the unit shift matrix with a circulant matrix, the spectral density of which in the limit $N\to \infty$ covers uniformly the unit circle and the matrix itself is normal, therefore its Green's function is the same as for the Haar unitary $\bb{U}$. This object was calculated in \cite{JAROSZGOERLICH,JAROSZUNIT} and reads
\begin{equation}
G_{\bb{U}}\left(\begin{array}{cc}
z & i\wb \\
iw & \zb 
\end{array}\right)=\left(\begin{array}{cc}
\frac{1}{2z}\left(1+\frac{|z|^2-|w|^2-1}{\sqrt{\alpha^2-4|z|^2}}\right) & -\frac{iw}{\sqrt{\alpha^2-4|z|^2}} \\
-\frac{i\wb}{\sqrt{\alpha^2-4|z|^2}} & \frac{1}{2\zb}\left(1+\frac{|z|^2-|w|^2-1}{\sqrt{\alpha^2-4|z|^2}}\right)
\end{array}\right),
\end{equation}
where $\alpha=|z|^2+|w|^2+1$. Endowed with this knowledge we can use the functional matrix equation \eqref{eq:SolutionSimplified}, obtaining a system of complex algebraic equations
\begin{eqnarray}
1&=&gz-v\wb+\frac{1}{2r}\left(1+\frac{r^2(|g|^2-|v|^2)-1}{\sqrt{\beta^2-4|g|^2r^2}}\right)+\frac{rv^2}{\sqrt{\beta^2-4|g|^2r^2}},\label{eq:unit1}\\
0&=&z\vb+\wb g+\frac{\vb}{2rg}\left(1+\frac{r^2(|g|^2-|v|^2)-1}{\sqrt{\beta^2-4|g|^2r^2}}\right)-\frac{r\gb v}{\sqrt{\beta^2-4|g|^2 r^2}},\label{eq:unit2}
\end{eqnarray}
with $\beta=r^2(|g|^2+|v|^2)+1$. The additional two equations from the $2\times 2$ matrix are the complex conjugate of the ones above. Nevertheless, in calculations it is convenient to treat $g$ and $\gb$ as independent variables and solve the extended system of 4 equations. The two additional equations assert that the solutions for $g$ and $\gb$ are mutually conjugated.

Having performed calculations which involve expansion into moments, one can put $w\to 0$ in further computations. It is clear that $v=0$ is a solution and  equation \eqref{eq:unit1} reduces to
\begin{equation}
1=gz+\frac{1}{2r}\left(1+\frac{r^2|g|^2-1}{\sqrt{(r^2|g|^2-1)^2}}\right).
\end{equation}
The square root yields an absolute value, which produces two holomorphic solutions
\begin{eqnarray}
g=\frac{1-r}{rz} & \mbox{for} & |z|\leq r-1, \\
g=\frac{1}{z} & \mbox{for} & |z| > r-1.
\end{eqnarray}

Bounds for the validity of these two solutions do not determine where the spectral density vanishes. In order to find the borderline of the spectrum, we need a non-holomorphic solution. To this end, we multiply \eqref{eq:unit2} by $g/\vb$ and subtract it from \eqref{eq:unit1}, obtaining
\begin{equation}
\frac{\vb}{v}=r\frac{|v|^2+|g|^2}{\sqrt{\beta^2-4|g|^2r^2}}.
\end{equation}
Since the r.h.s. is real, $v$ is either real or purely imaginary. The choice of the solution is equivalent to the choice of the $2\times 2$ matrix representation of a quaternion \eqref{eq:BlockStructure}. In our convention we take $v$ real and get the equation for $g$
\begin{equation}
4g^3r^3z^3+4g^2z^2r^2(1-r)+gzr\left((1-r)^2-|z|^2\right)-|z|^2=0.
\end{equation}  
The structure of the equation above suggests to substitute $f=gz$, which brings it to the following form
\begin{equation}
4f^3r^3+4f^2r^2(1-r)+fr\left((1-r)^2-|z|^2\right)-|z|^2=0. \label{eq:unitCDF}
\end{equation}
It is clear that $f$ depends only on $|z|$, which means that this is a radial cumulative distribution function for matrix ensembles, the spectrum of which is rotationally invariant on the complex plane \cite{FEINBERGZEE}. The spectral density is then given by
\begin{equation}
\rho(z,\zb)=\frac{1}{\pi}\partial_{\zb}g(z,\zb)=\frac{1}{2\pi |z|}f'(|z|).
\end{equation}
Moreover, the eigenvector correlator \eqref{eq:Correlator}, calculated from the system of equations \eqref{eq:unit1} \eqref{eq:unit2} inside the spectrum reads
\begin{equation}
O(|z|)=\frac{1}{\pi}\frac{1}{2fr^2+r-r^2}-\frac{f^2}{\pi |z|^2},
\end{equation}
 outside the spectrum the correlator vanishes.

It is worth reminding that the obtained spectrum of a sandwiched matrix differs from the one of a product of the anti-Wishart matrix and a Haar unitary by an absence of $T-N$ zero modes. Indeed, a transformation $F=(1-r)+fr$ brings \eqref{eq:unitCDF} to \eqref{CDFequation}. These two matrices, obtained by a product of 3 matrices but in a different order, share the same spectral properties, which is a consequence of the Sylvester's determinant identity. However, the block trace operation is not cyclic, therefore the eigenvector correlator distinguishes between these ensembles. This correlator for the product of anti-Wishart with Haar unitary, which falls into the range of applicability of the Haagerup-Larsen theorem, can be calculated from a simple formula \eqref{eq:HaagLarsCorr}.

To calculate the  boundary of the spectrum we exploit the fact that $f$ is a radial cumulative distribution function and $f=1$ corresponds to the outer radius. If $N<T$ there are no zero modes, and the inner spectral radius is determined by the condition $f=0$, while for $N>T$ the presence of zero eigenvalues requires that $f=1-1/r$ gives the inner radius. Their values agree exactly with \eqref{eq:spectralradii}, thus we can finally write the spectral density
\begin{equation}
\rho(z,\zb)=\max\left(\frac{r-1}{r},0\right) \delta^{(2)}(z)+\frac{1}{2\pi |z|}f'(|z|)\theta(|z|-s_{int})\theta(s_{ext}-|z|),
\end{equation}
where $f$ is a real valued branch of the solution of \eqref{eq:unitCDF} and $s_{int}, s_{ext}$ are given by \eqref{eq:spectralradii}. 

\subsection{Going deeper with the lag}

Considering a deeper lag, one introduces an additional time scale to the problem, namely, the depth of the lag $\tau$. In the standard limiting procedure $N,T\to\infty$, keeping $\tau$ constant does not produce new results. The spectral density and the eigenvector correlator tend to the already solved case of the unit time shift. This model works well if the time lag is much smaller than the length of time series.

In real situations time lags sometimes can be comparable with the temporal length of the data. In order to capture this property, we perform  the limiting procedure when all $N,T,\tau$ tend to infinity with $r=N/T$ and additionally $\beta=\tau/T$ fixed. For the convenience we introduce another parameter $\alpha=(1-\tau/T)^{-1}$ which is finite in this limit.

The algebraic equation for the generalized Green's function of the lagged correlation matrix in this double scaling limit can be easily obtained from \eqref{eq:Solution}, by taking into account that $\frac{1}{T-\tau}\bb{D}=\frac{1}{T}\alpha \bb{D}$ and it takes the following form
\begin{equation}
\left[Q-\frac{\alpha}{T}\btr\left({\cal D}[\idm_{2T}-\alpha r(G\otimes\idm_{T}){\cal D}]^{-1}\right)\right]G=\idm_{2}, \label{eq:algebraicLAg}
\end{equation} 
with $D_{ts}=\delta_{t+\tau,s}$ and
\begin{equation}
{\cal D}=\left(\begin{array}{cc}
\bb{D} & 0 \\
0 & \bb{D}^{T}
\end{array}\right).
\end{equation}

This equation simplifies considerably if $\tau/T$ is a fraction with a small denominator. The delay matrix can then be written as a Kronecker product of a low dimensional matrix and the identity matrix. Here, we present a step-by-step derivation for the simplest instance $T=2\tau$. It is convenient to write matrices as follows
\begin{equation}
{\cal D}=\left(\begin{array}{cccc}
0 & 1 & 0 & 0 \\
0 & 0 & 0 & 0 \\
0 & 0 & 0 & 0 \\
0 & 0 & 1 & 0 
\end{array}\right)\otimes \idm_{\tau},\, G\otimes\idm_{T}=\left(\begin{array}{cccc}
g & 0 & i\vb & 0 \\
0 & g & 0 & i\vb \\
iv & 0 & \gb & 0 \\
0 & iv & 0 & \gb
\end{array}\right)\otimes \idm_{\tau}, \, \idm_{2T}=\idm_{4}\otimes\idm_{\tau},
\end{equation}
which reduces the matrix within the block trace in \eqref{eq:algebraicLAg} to a Kronecker product of a $4\times 4$ matrix with the identity matrix, simplifying considerably the matrix inversion and the block trace operation. Upon setting $w=0$, one obtains a system of complex equations
\begin{equation}
\frac{2r|v|^2}{1+4r^2|v|^2}+gz=1,\qquad \vb\left(\frac{2r\gb}{1+4r^2|v|^2}-z\right)=0.
\end{equation}
Once again $v=0$ together with $g=1/z$ is a solution valid for large $|z|$. Assuming $v\neq 0$, we eliminate $|v|^2$, obtaining the quadratic equation for $f=gz$
\begin{equation}
4f^2r^2+2fr(1-2r)-|z|^2=0.
\end{equation}
Substitution $f=1$ yields the spectral radius $s_{ext}=\sqrt{2r}$. Taking appropriate branch of solutions and differentiating with respect to $|z|$, one obtains the spectral density
\begin{equation}
\rho(z,\zb)=\frac{1}{2\pi r \sqrt{(1-2r)^2+4|z|^2}} \theta(\sqrt{2r}-|z|).
\end{equation}
The eigenvector correlator in this case is given by the formula
\begin{equation}
O(z,\zb)=\frac{2r-1-2|z|^2+\sqrt{(2r-1)^2+4|z|^2}}{8\pi r^2|z|^2}\theta(\sqrt{2r}-|z|).
\end{equation}

The spectral density is the same as the one for a product of two independent rectangular Wishart matrices \cite{BJLNSproducts}. We argue here that this is the case for all $\tau\geq T/2$. Let us write matrices in the block form
\begin{eqnarray}
\bb{X}=\left(\bb{A}_{N\times\tau},\bb{B}_{N\times(T-\tau)}\right),\quad \bb{D}(\tau)=\left(\begin{array}{cc}
0_{\tau\times(T-\tau)} & \idm_{\tau\times\tau} \\
0_{(T-\tau)\times(T-\tau)} & 0_{(T-\tau)\times\tau}
\end{array}\right), \\
 \bb{X}^{\dagger}=\left( \begin{array}{c}
\bb{F}_{(T-\tau)\times N} \\
\bb{H}_{\tau\times N}
\end{array}\right).
\end{eqnarray}

In this representation the lagged correlation matrix takes the form $\bb{C}=\frac{1}{T-\tau}\bb{AH}$. For $\tau\geq T/2$ blocks $\bb{A}$ and $\bb{H}$ have no common elements, and the lagged correlation matrix reduces to the product of two independent  rectangular matrices of size $N\times \tau$ multiplied by a factor $(T-\tau)^{-1}$. A RMT benchmark for the lagged correlation matrix as a product of two independent rectangular Gaussian matrices, as proposed by Livan \cite{LIVAN}, turns out to be valid for $\tau\geq T/2$.

\begin{figure}
\includegraphics[width=0.49\textwidth]{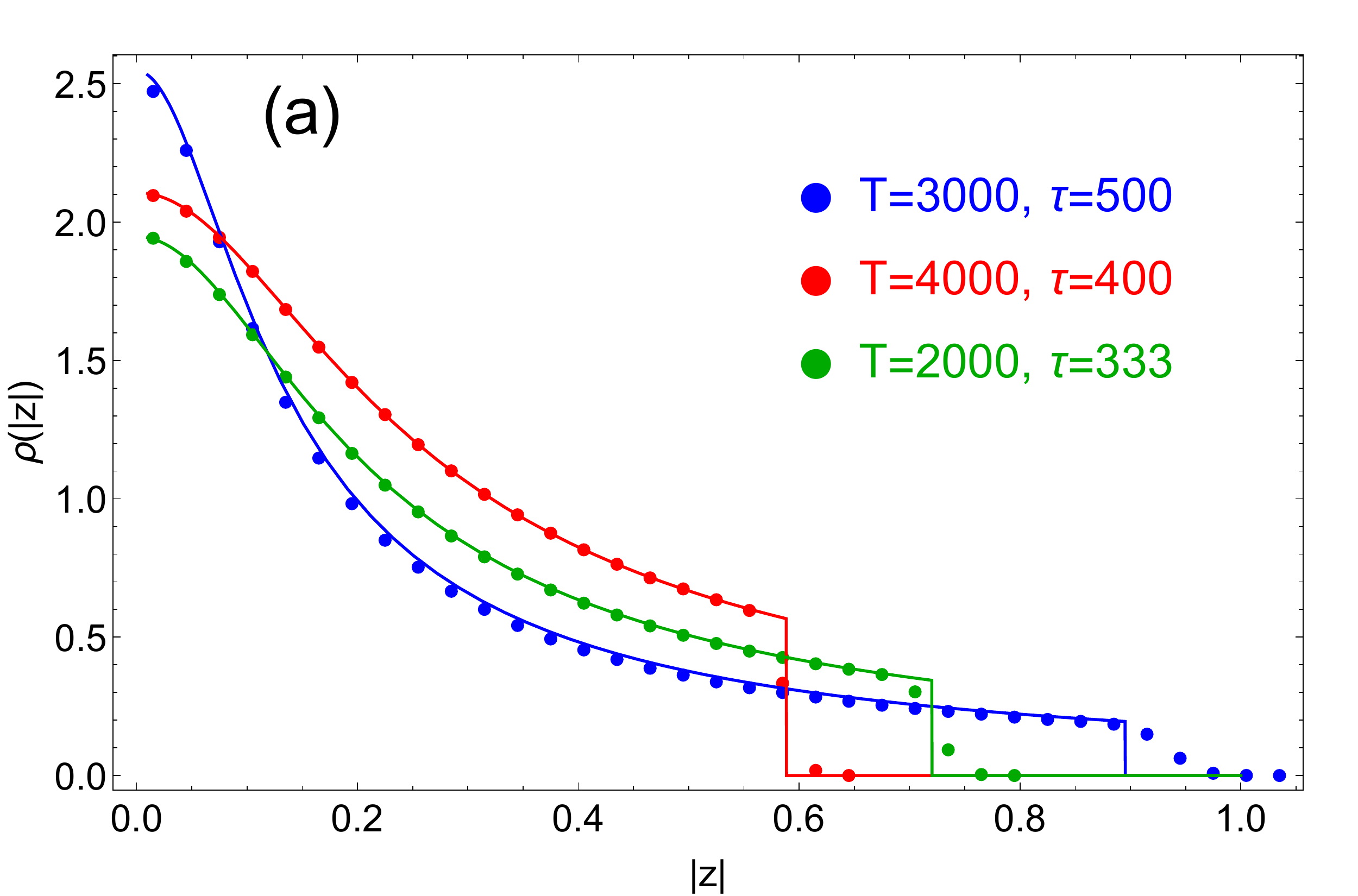} \includegraphics[width=0.49\textwidth]{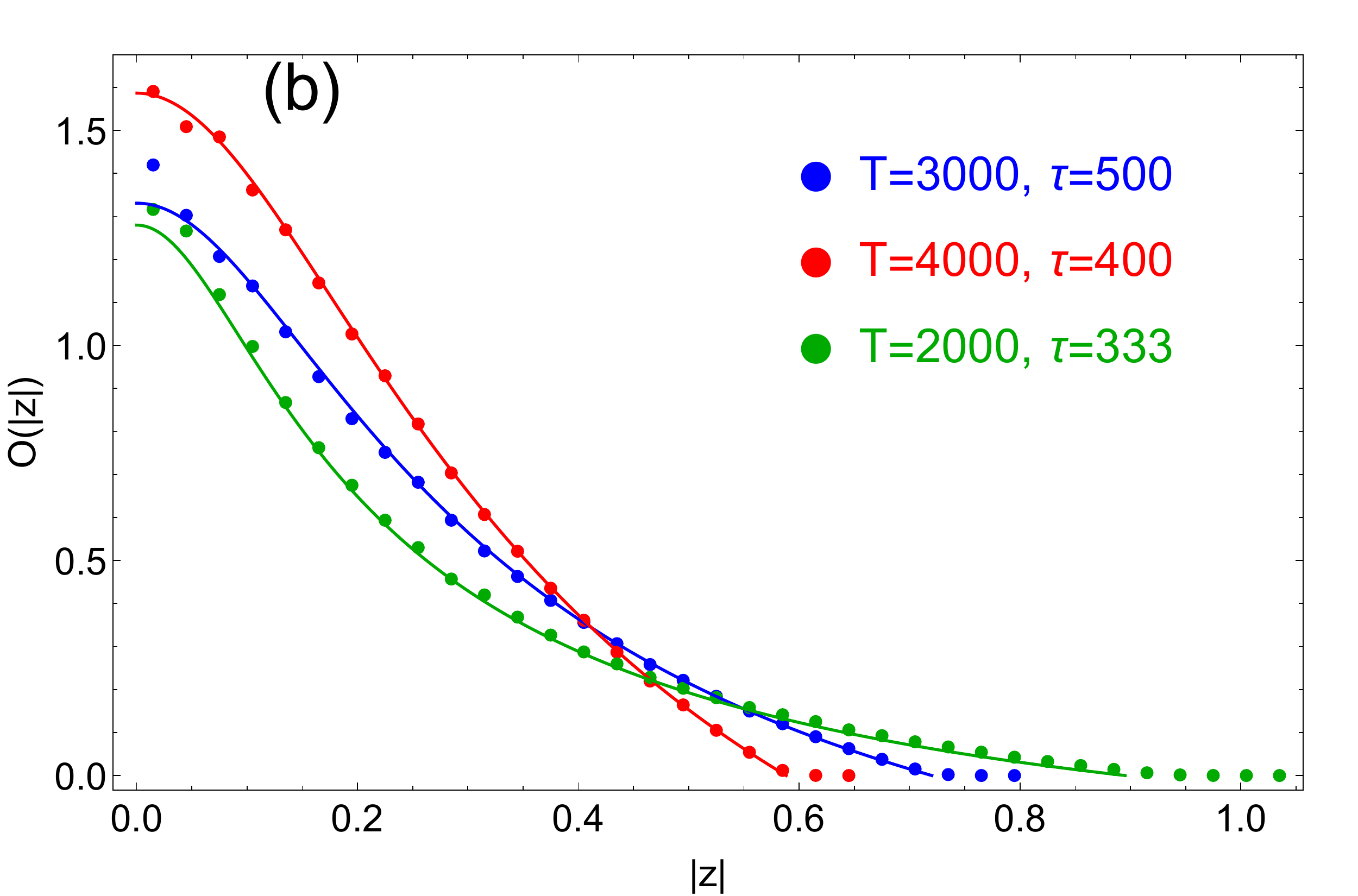}
\caption{Radial spectral densities (a) and the eigenvector correlator (b) obtained by the diagonalization of $4000$ Gaussian lagged correlation matrices of size $N=1000$ with various rectagularities and lag depths (dots) juxtapposed with the numerical solutions of the algebraic equation \eqref{eq:algebraicLAg}. The discrepancies near the edges of the spectrum are the effects of the finite size of matrices.
\label{fig:VariousLags}}
\end{figure}

Considering various lags such that the denominator of $\tau/T$  is greater or equal to 3, one obtains a system of coupled polynomial equations for $f=gz$ and $|v|^2$, which cannot be solved analytically. Nevertheless, the numerical solutions agree with Monte Carlo simulations, as presented in Fig. \ref{fig:VariousLags}.

\begin{figure}
\begin{center}
\includegraphics[width=0.7\textwidth]{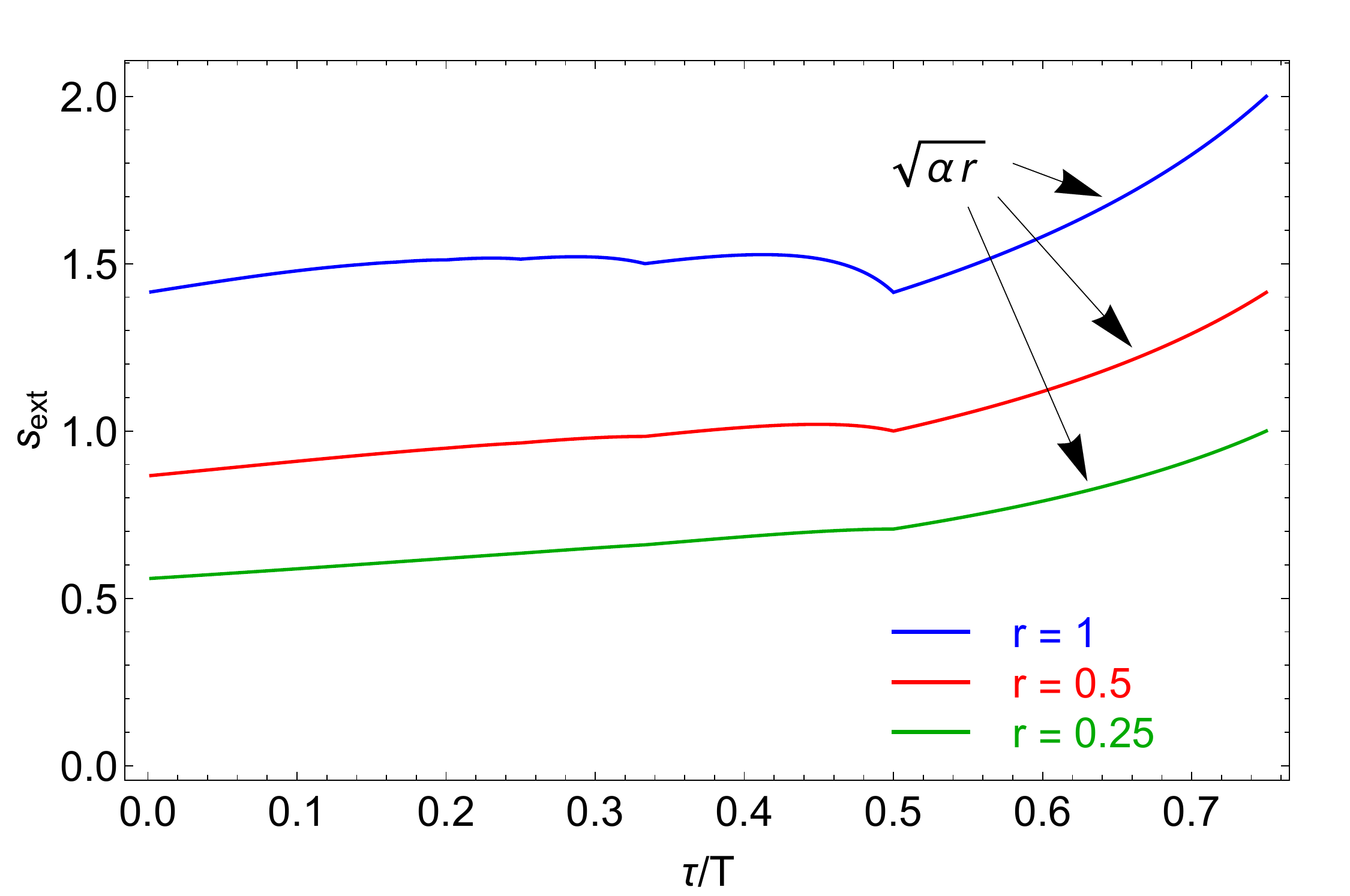} \end{center}
\caption{Dependence of the spectral radius as a function of the lag depth for 3 arbitrary rectangularities. For $\tau>T/2$ there is a universal behavior  $s_{ext}=\sqrt{\alpha r}$. The cusps are consequences of the discontinuity of the ceiling function.
\label{fig:SpecRad}}
\end{figure}

Even more important from the perspective of data analysts is the radius of a circle bounding the support of the spectral density. By imposing matching conditions of the holomorphic and non-holomorphic solutions for $g$ we obtain an implicit equation for the external radius $s_{ext}$
\begin{equation}
\sum_{k=1}^{M-1}\left(\frac{\alpha r}{s_{ext}}\right)^{2k}(1-k\beta)=r, \label{eq:SpecRad}
\end{equation}
where $\beta=\tau/T$, $M=\left\lceil \frac{T}{\tau} \right\rceil$ and $\left\lceil x\right\rceil$ denoted the ceiling function. Details of the derivation are relegated to Appendix \ref{sec:SpectralRadius}, plots depicting the solutions are presented in Fig. \ref{fig:SpecRad}. In the limit $\tau/T\to 0$ equation \eqref{eq:SpecRad} yields $s_{ext}=\sqrt{r(r+1)}$, reproducing the result for the unit time lag.

\section{Conclusions } \label{sec:Conclusions}

In this paper we have analyzed a broad class of large lagged correlation matrices, using the tools of random matrix theory. 
We started from the comparison of several  previously studied cases (symmetrization, Abelization, whitening)  using the powerful tools of free random variables. This approach has not only allowed us to rederive  the already existing results, but also to resolve some controversies in the literature. Then, we considered the non-Hermitian problem of the lagged correlation matrix using the concise description of the diagrammatic techniques in terms of the so-called generalized Green's functions (quaternionic Green's function).  Two  new results include: (i) the algebraic equation for the quaternionic resolvent \eqref{eq:Solution} and (ii) the solution  for deeper time lags - in particular the implicit expression for the external radius of the complex spectrum \eqref{eq:SpecRad}. The solution for the lagged correlation matrix is obtained as a particular instance of a more general random matrix problem, namely the non-Hermitian generalization of the Wishart ensemble.
Additionally, we have addressed the issue of the eigenvector correlators in such matrices. To the best of our knowledge, such objects were not studied so far in non-Hermitian generalizations of multivariate statistics, perhaps due to the lack of the pertinent mathematical methods. We hope that our framework will fill this gap. 
 
 Last but no least, several generalizations of the presented formalism  are possible.  The  first is the inclusion of spatio-temporal correlations for time-lagged covariances, which will cause the spectra to lose the rotational symmetry. An important step in this direction was already done~\cite{JAROSZLAG}, but resulting formalism is so far quite complicated. The other  direction  is the departure from Gaussianity in favor of heavy tailed data, so abundant in real-world complex systems. 
 
The intention of the paper was also to be self-consistent and  to promote new techniques for large lagged matrices, therefore we decided to present the detailed calculations and to add extensive appendices.  We also successfully confronted  formulae obtained with numerical, synthetic data. 
The presented formalism can be applied  to the empirical analysis in any field where lagged correlations are essential - we have already outlined the domains of science in the introduction. From our point of view, the most interesting are the  hidden correlations in  biomedical data and we will address this issue in the sequel to this work~\cite{CHIALVO}. 

\subsubsection*{Acknowledgments}
We thank  Zdzis\l aw Burda, Romuald Janik  and Andrzej Jarosz for discussions and for the check of some of the calculations at the early stage of this work. This work was supported  by the Grant DEC-2011/02/A/ST1/00119 of the National Center of Science.  
WT appreciates also the support from Polish Ministry of Science and Higher Education through the Diamond Grant 0225/DIA/2015/44 and the scholarship of Marian Smoluchowski Research Consortium Matter Energy Future from
KNOW funding.

\appendix

\section{Product of two free projectors} \label{App:FreeProjectors}
We present here the free random variables calculation of the spectral density for the product of two identical but mutually free projection operators. 
We start from the definition of the moment generating function $M_\bb{P}(z)=zG_\bb{P}(z) -1$, where $G_\bb{P}(z)= \alpha \frac{1}{z-1} +(1-\alpha)\frac{1}{z}$,  which  corresponds 
to $\alpha=N/T$  fraction of eigenvalues 1 and $(1-\alpha)$ fraction of eigenvalues 0. Explicit calculation gives $M_\bb{P}(z)=\frac{\alpha}{z-1}$. From the definition of the functional inverse of the generating function, $M_\bb{P}[N_\bb{P}(z)]=z$, we read $N_\bb{P}(z)=\frac{z+\alpha}{z}$. 
The multiplication law reads
\be
N_{1\cdot 2}(z)=\frac{z}{1+z} N_1(z) \cdot N_2(z),
\ee
so the N-transform for the product of identical projections reads 
\be
N_{\bb{P}^2}(z)=\frac{(z+\alpha)^2}{z(z+1)}.
\ee
Using again the definition of the functional inverse, we arrive at the quadratic equation for $M_{\bb{P}^2}(z)$
\be
(z-1)M_{\bb{P}^2}^2(z) +(z-2\alpha)M_{\bb{P}^2}(z) -\alpha^2 =0.
\label{quadraticM}
\ee
The Green's function therefore reads  (note $M_{\bb{P}^2}(z)=zG_{\bb{P}^2}(z)-1  $)  
\be
G_{\bb{P}^2}(z)=\frac{1}{2z(z-1)} \left[z-2(1-\alpha)  +\sqrt{z(z-4\alpha(1-\alpha))}\right],
\ee
where the  have chosen the solution of the quadratic equation (\ref{quadraticM}) reproducing the asymptotic behavior  $G_{\bb{P}^2}(z) \sim 1/z$ for large $|z|$.  
Taking the imaginary part  we finally arrive at the spectral  density
\be
\rho_{P^2}(\lambda)=(1-\alpha)\delta(\lambda) +{\rm max}(2\alpha-1,0) \delta(\lambda-1) +
\frac{\sqrt{4\alpha(1-\alpha)-\lambda}}{2 \pi \sqrt{\lambda}(1-\lambda)},
\label{finalidentJac}
\ee
where the continuous  part of the spectrum spans the interval $[0, 4\alpha(1-\alpha)]$. 
This is a special case of the so-called free Jacobi distribution, obtained first time for two different, free  projectors in ~\cite{VDN92} ({\it cf.}  Example 3.6.7)  and then studied e.g. in ~\cite{BOUCHAUD,COLLINS}. 

\section{Details of the diagrammatic expansion} \label{sec:Diagrams}

\begin{figure}
\begin{center}
\includegraphics[width=0.95\textwidth]{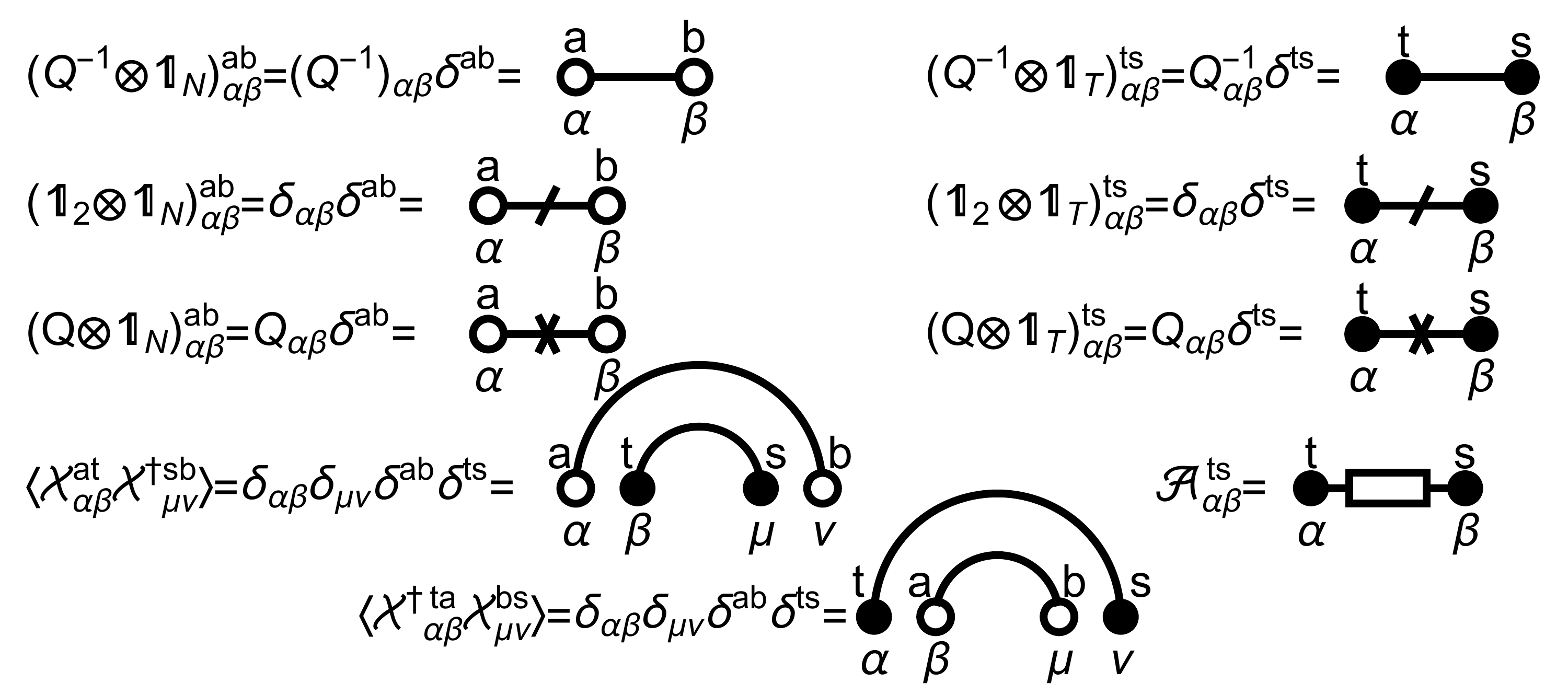}
\end{center}
\caption{
A pictorial representation of the expressions in the moment expansion of the quaternionic Green's function.
\label{Fig:DiagrammaticRules}
}
\end{figure}

Our aim is to calculate the quaternionic resolvent expanded into moments \eqref{eq:Gexpansion}. Wick's theorem allows one to express higher moments of $\cX$ via its second cumulants only. The price that we pay is the increase of the number of terms at each order in the sum. The corresponding expressions become lenghty and the multitude of symbols obscures the internal structure of the expressions. To get rid of letters, we represent each term in the expansion by a diagram. All pictures composing diagrams and expressions corresponding to them are presented in Fig. \ref{Fig:DiagrammaticRules}.

Each diagram has two endpoints denoted as dots, they correspond to matrix elements that are indexed by two letters. Greek letters written beneath the circle refer to a quaternion space and run from 1 to 2. Latin letters from the beginning of the alphabet placed over empty circles correspond to the matrix space and run from $1$ to $N$. To distinguish matrix elements, the range of which is from $1$ to $T$, we draw filled circles and use Latin letters from the middle of the alphabet. The expectation of two matrix elements is represented as a double arc joining particular elements. Furthermore, the sandwiched deterministic matrix is depicted as a rectangle.

A sum over repeating indices is denoted as merged circles. For clarity, we do not write indices on diagrams explicitly. The flow of the quaternion indices is simply from the l.h.s. to the the r.h.s of the diagram, while the flow of the matrix indices follows the direction of lines and arcs.

\begin{figure}
\includegraphics[width=\textwidth]{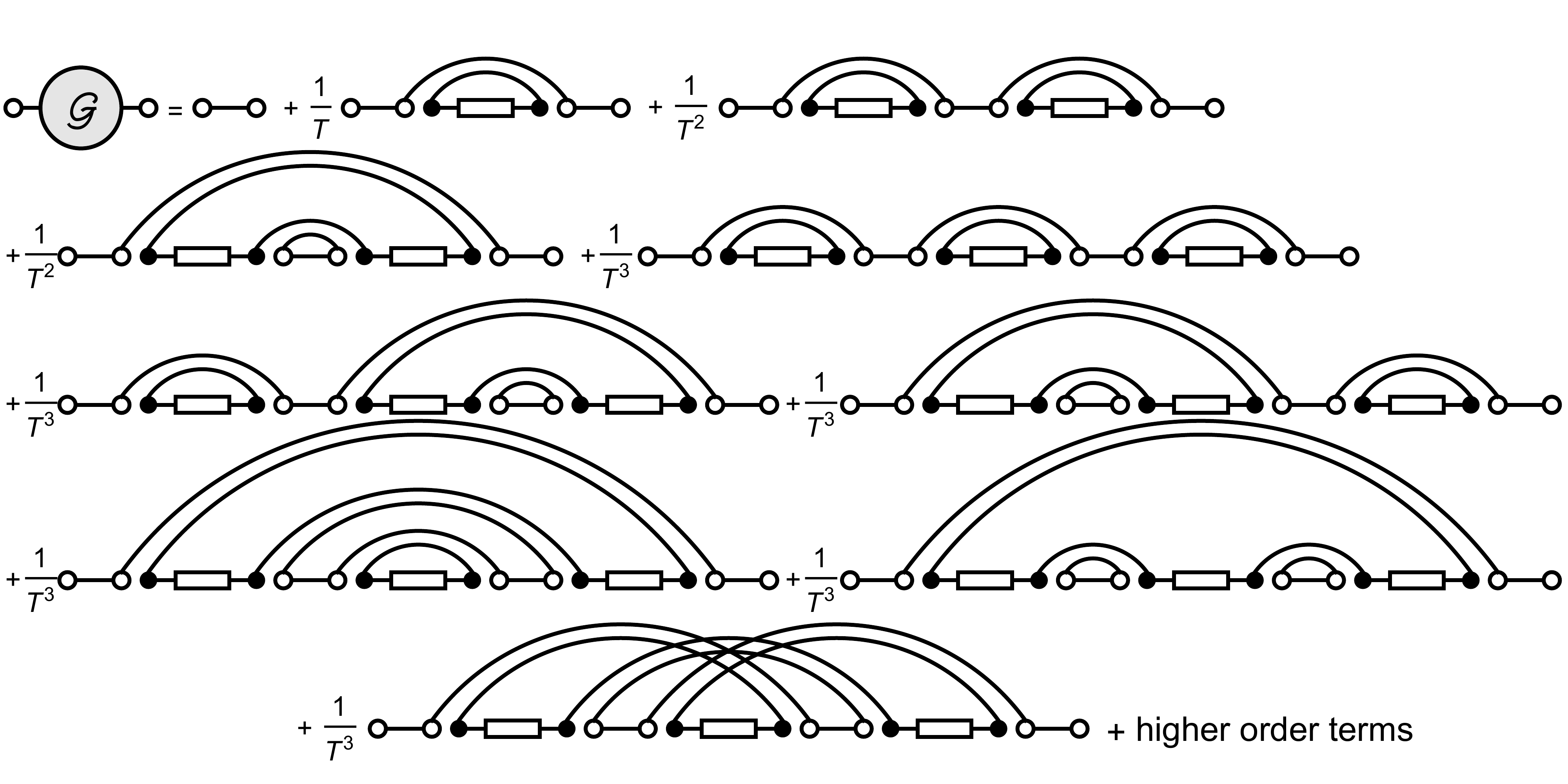}
\caption{
All diagrams up to the third order contributing to the Green's function. The last diagram is not planar and its contribution vanishes in the $N,T\to\infty$ limit.
\label{Fig:DiagrGreensFunction}}
\end{figure}

In the expansion of the quaternionic resolvent $\cG$ all matrix elements have to be paired and each pair contains one elements from $\cX$ and one from $\cX^{\dagger}$. Since the Wick's theorem produces a sum over all possible contractions, the $n$-th term in the expansion is represented by $n!$ distinct diagrams. All diagrams from the first four terms are presented in Fig. \ref{Fig:DiagrGreensFunction}.

Direct calculations lead to the observation that every loop in the diagrams gives a factor $T$ if it contains filled circles and $N$ for empty ones. A careful insight into the structure of the diagrams leads to the conclusion that in the limit $T,N\to\infty$ with $N/T$ fixed only planar diagrams contribute.

\begin{figure}
\includegraphics[width=\textwidth]{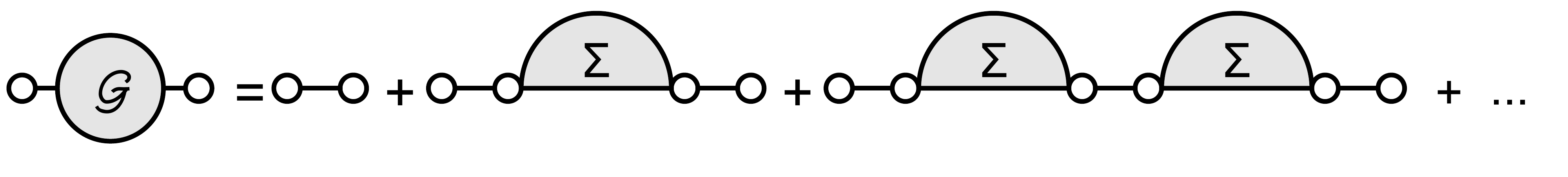}
\caption{Diagrammatic representation of the Schwinger-Dyson equation.
\label{Fig:SD1}}
\end{figure}

We distinguish a particular class of diagrams that cannot be split into two by a single cut of an internal horizontal line corresponding to $\cQ^{-1}$ and call them one-line irreducible (1LI). We denote $\Sigma$ as the sum of all 1LI diagrams with amputated external legs. By amputating we mean dividing the corresponding expression (here $\cQ^{-1}$) out. $\Sigma$, which we also refer to it as self-energy, turns out to be a building block of the quaternionic resolvent. One can rearrange all terms contributing to $\cG$ into groups such that the first one consists of 1LI diagrams, the second group contains diagrams that can be split by a single cut into two 1LI diagrams, the third is comprised of diagrams that have two such vulnerable lines, and so on. This expansion is presented in Fig. \ref{Fig:SD1}. In the index notation, this equation, also known as Schwinger-Dyson equation, reads
\begin{equation}
\cG_{\alpha\beta}^{ab}=(\cQ^{-1})^{ab}_{\alpha\beta}+(\cQ^{-1})^{ac}_{\alpha\gamma}\Sigma_{\gamma\varphi}^{cd}(\cQ^{-1})^{db}_{\varphi\beta}+(\cQ^{-1})^{ac}_{\alpha\gamma}\Sigma_{\gamma\varphi}^{cd}(\cQ^{-1})^{de}_{\varphi\rho}\Sigma_{\rho\mu}^{ef}(\cQ^{-1})^{fb}_{\mu\beta}+\ldots .
\end{equation}
This is a geometric series, which one can sum and write in a closed form
\begin{equation}
\left(\cQ^{ac}_{\alpha \gamma}-\Sigma^{ac}_{\alpha \gamma}\right)\cG^{cb}_{\gamma\beta}=\delta^{ab}\delta_{\alpha\beta}. \label{eq:sd1}
\end{equation}

\begin{figure}
\includegraphics[width=\textwidth]{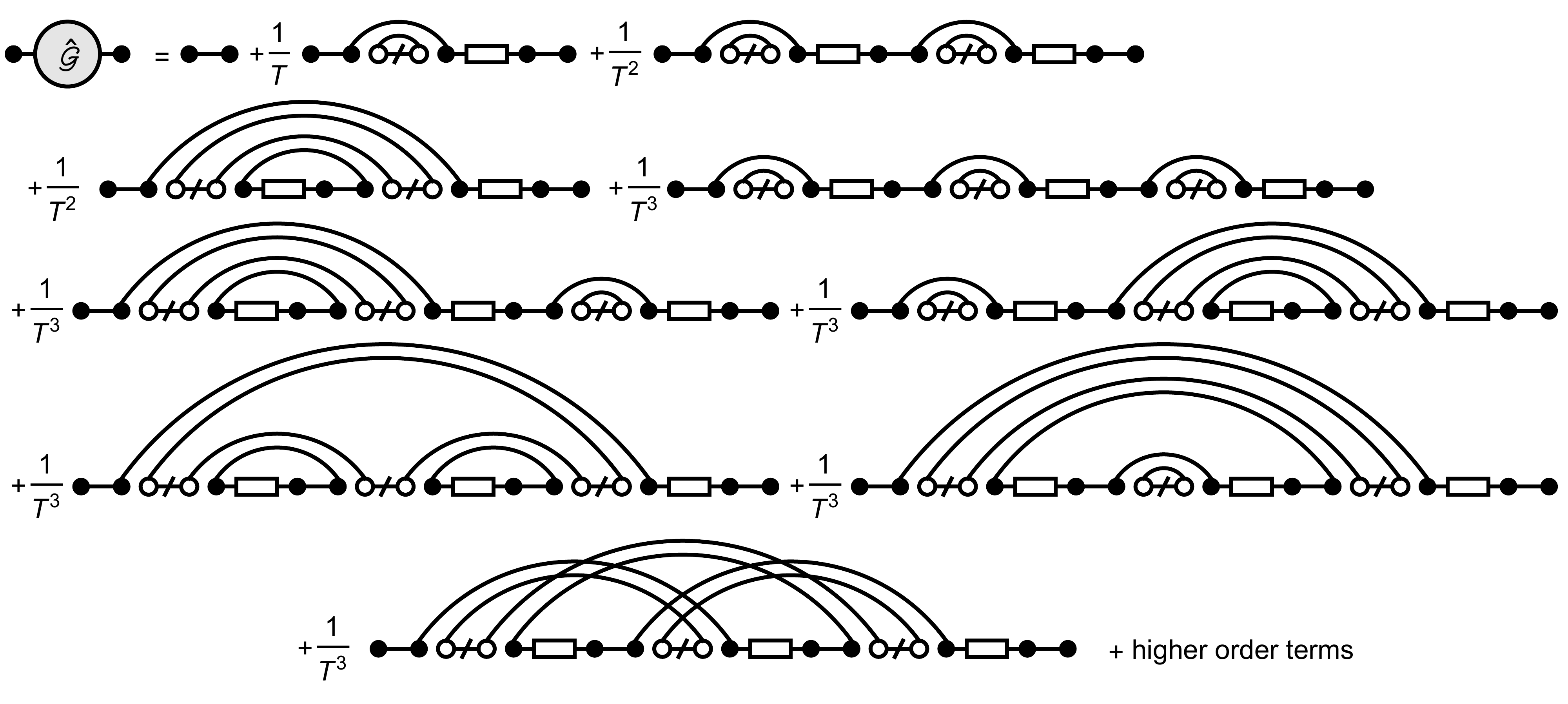}
\caption{Feynman diagrams contributing to the auxiliary Green's function up to fourth order.
\label{Fig:DiagrGreensDual}}
\end{figure}

We introduce an auxiliary expansion of a $T\times T$ matrix $\hat{\cY}=\frac{1}{T}\cX^{\dagger}\cX\cA$, which is a cyclic permutation of matrices forming $\cY$. We denote all quantities pertaining to this auxiliary expansion with the same symbols as for the main problem, but we put a hat over them. We consider the  quaternionic resolvent associated with this matrix
\begin{equation}
\hat{\cG}=\<(Q\otimes\idm_{T}-\hat{\cY} )^{-1}\>=\cQ^{-1}+\frac{1}{T}\<\cQ^{-1}\cX^{\dagger}\cX\cA\cQ^{-1}\>+\ldots  \,.
\end{equation}
The diagrammatic representation of expressions generated by the first four terms of its expansion is presented in Fig. \ref{Fig:DiagrGreensDual}. There still holds a relation between the quaternionic resolvent and 1LI diagrams summed to the self-energy $\hat{\Sigma}$
\begin{equation}
\left(\cQ_{\alpha \gamma}^{tr}-\hat{\Sigma}_{\alpha \gamma}^{tr}\right)\hat{\cG}^{rs}_{\gamma\beta}=\delta^{ts}\delta_{\alpha \beta}. \label{eq:sd2}
\end{equation}

\begin{figure}
\includegraphics[width=\textwidth]{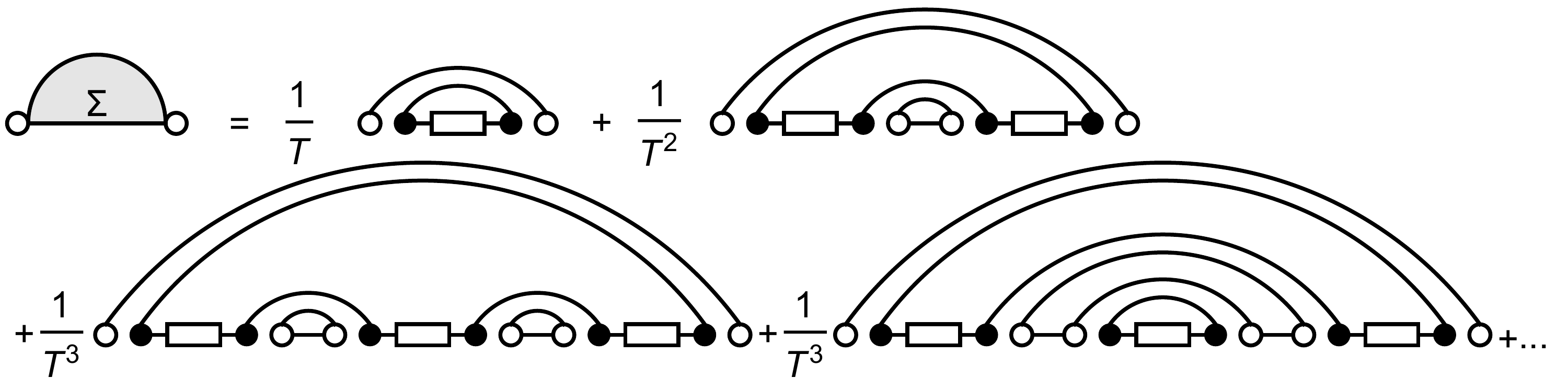}
\caption{
Four the simplest 1LI diagrams corresponding to the self-energy.
\label{Fig:DiagrSigma}}
\end{figure}

\begin{figure}
\includegraphics[width=\textwidth]{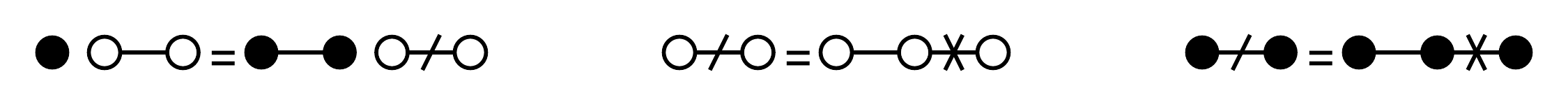}
\caption{Simple diagrammatic identities which allow us to transfigurate diagrams.
\label{Fig:DiagrIdentities}}
\end{figure}

\begin{figure}
\includegraphics[width=\textwidth]{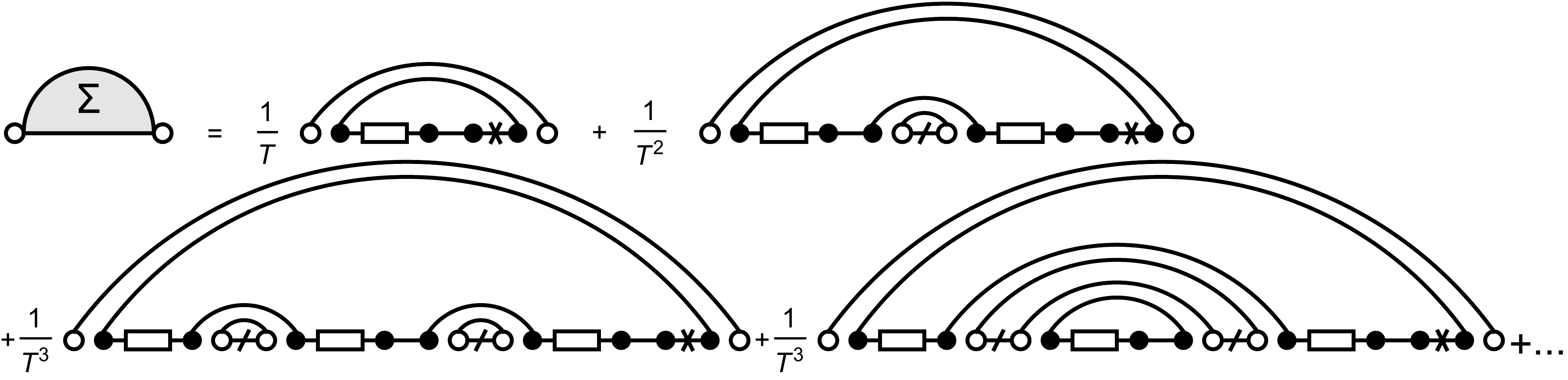}
\caption{Transfigurated 1LI diagrams contributing to the Green's function.
\label{Fig:DiagrSigma2}}
\end{figure}

\begin{figure}
\includegraphics[width=0.49\textwidth]{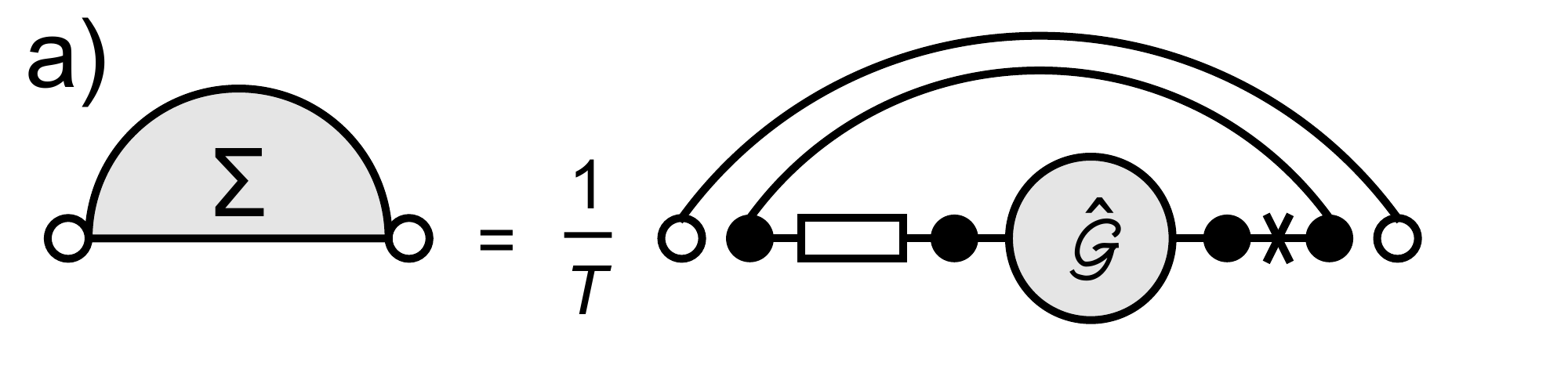} \includegraphics[width=0.49\textwidth]{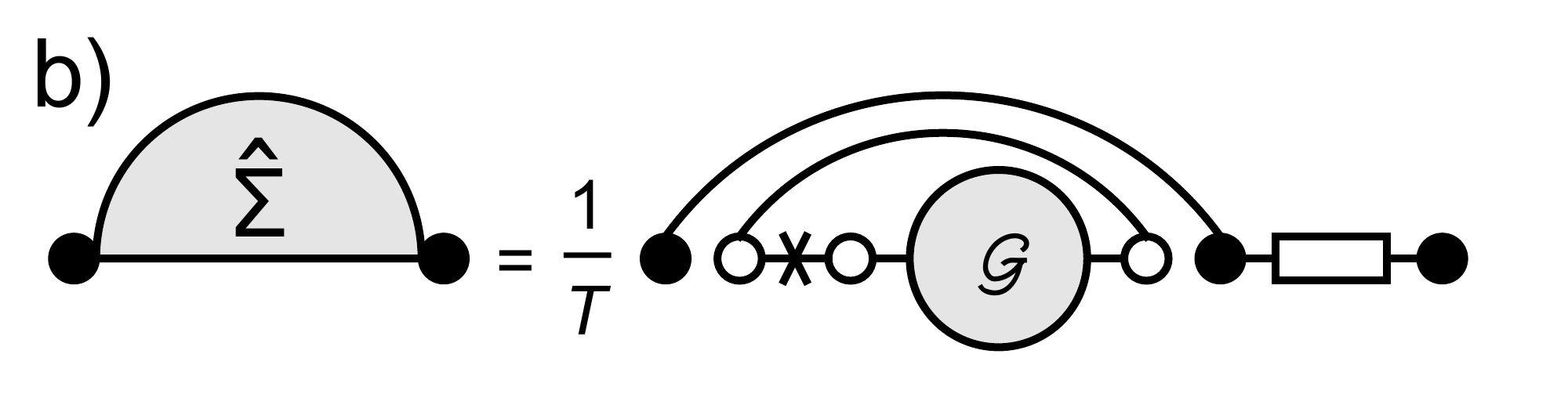} 
\caption{Relations between the Green's function and 1LI diagrams of the dual problem (a) and the auxiliary Green's function and 1LI diagrams of the original matrix problem (b). \label{Fig:DiagrSol}}
\end{figure}

First four diagrams contributing to the self-energy $\Sigma$ are presented in Fig. \ref{Fig:DiagrSigma}. By 
simple diagrammatic identities from Fig. \ref{Fig:DiagrIdentities} we bring the diagrams to the form which is depicted in Fig. \ref{Fig:DiagrSigma2}. The structure of 1LI diagrams is now evident and one can relate the self-energy and the quaternionic resolvent of the auxiliary problem, as presented in Fig. \ref{Fig:DiagrSol}a. Analogous transfiguration of 1LI diagrams of the auxiliary problem leads to a simple diagrammatic relation between the self-energy of the auxiliary problem and the desired quaternionic resolvent, see Fig. \ref{Fig:DiagrSol}b.

These relation in the index notation read
\begin{eqnarray}
\Sigma_{\alpha\beta}^{ab}&=&\frac{1}{T}\delta^{ab}\cA^{ts}_{\alpha\gamma}\hat{\cG}^{st}_{\gamma\mu}Q_{\mu\beta}, \\
\hat{\Sigma}_{\alpha \beta}^{st}&=&\frac{1}{T}Q_{\alpha \gamma}\cG^{aa}_{\gamma\mu}\cA^{st}_{\mu\beta}.
\end{eqnarray}
The above equations together with the Schwinger-Dyson equations \eqref{eq:sd1}, \eqref{eq:sd2} form a closed system that determines the solution for the quaternionic Green's function of the matrix $\bb{Y}$.

\section{Spectral radius of the lagged correlation matrix} \label{sec:SpectralRadius}
Representing matrices under the block trace as $T\times T$ blocks and making use of the block inverse formula
\begin{equation}
\left(\begin{array}{cc}
\bb{A} & \bb{B} \\
\bb{C} & \bb{F}
\end{array}\right)^{-1}=\left(\begin{array}{cc}
(\bb{A}-\bb{BF}^{-1}\bb{C})^{-1} & -\bb{A}^{-1}\bb{B}(\bb{F}-\bb{CA}^{-1}\bb{B})^{-1} \\
-\bb{F}^{-1}\bb{C}(\bb{A}-\bb{BF}^{-1}\bb{C})^{-1} & (\bb{F}-\bb{CA}^{-1}\bb{B})^{-1}
\end{array}\right),
\end{equation}
one can analyze equation \eqref{eq:algebraicLAg} in the most general form. Considering now the delay matrix $D_{t,t'}=\delta_{t+\tau,t'}$ in the sandwich, we obtain the expression for the block trace in \eqref{eq:algebraicLAg} 
\begin{equation}
\frac{1}{T-\tau}\btr \left(\cA [\idm_{2T}-\alpha r  (G\otimes \idm_{T}) \cA ]^{-1}\right)=\left(\begin{array}{cc}
F & ir \alpha \vb H \\
ir\alpha v J & K
\end{array}\right),
\end{equation}
with $F=\bar{K}$ and $H=\bar{J}$ given by
\begin{eqnarray}
F=\frac{1}{T-\tau}\Tr\left(\bb{D}\left[\idm_{T}-\alpha rg \bb{D}+r^2\alpha^{2}|v|^2 \bb{D}^{T}(\idm_{T}-\alpha r \gb \bb{D}^T)^{-1}\bb{D}\right]^{-1}\right), \\
H=\frac{1}{T-\tau}\Tr\left(\bb{D}[\idm_{T}-\alpha rg \bb{D}]^{-1}\bb{D}^{T}\left[\idm_{T}-\alpha r \gb \bb{D}^{T}+ \right.\right. \nonumber \\ \left.\left. r^2\alpha^2|v|^2\bb{D}(\idm_{T}-r\alpha g \bb{D})^{-1}\bb{D}^{T}\right]^{-1}\right).
\end{eqnarray}
Matrix equation \eqref{eq:algebraicLAg} produces two scalar complex equations
\begin{eqnarray}
(z-F)g+|v|^2r\alpha H =1, \label{eq:SpRad1} \\
 v( r \alpha g J-\zb+K)=0. \label{eq:SpRad2}
\end{eqnarray}
Again, after simple computations $v=0$ and $g=z^{-1}$ turn out to be solutions of this system, describing the Green's function on the exterior of the spectrum. Assuming now $v\neq 0$, we divide \eqref{eq:SpRad2} by $v$. The resulting system of equations determines the Green's function within the support of the spectral density. This system is in general very complicated, but, regardless of its solution, the holomorhic and non-holomorphic solutions for $g$ must match exactly at the boundary of the spectrum. Upon substitution $g=1/z$ and $v=0$ \eqref{eq:SpRad1} is trivially satisfied, while \eqref{eq:SpRad2} yields the equation for $z$ and $\zb$
\begin{eqnarray}
\frac{r\alpha}{T-\tau} \Tr\left(\bb{D}^{T}\left[\idm_{T}-\frac{\alpha r}{\zb}\bb{D}^T\right]^{-1} \bb{D}\left[\idm_{T}-\frac{\alpha r}{z}\bb{D}\right]^{-1}\right)= \nonumber \\ |z|^2-\frac{z}{T-\tau}\Tr\left( [\idm_T-\frac{\alpha r}{\zb} \bb{D}^T]^{-1}\bb{D}^{T}\right).
\end{eqnarray}
We expand $[1-\frac{\alpha r}{z}\bb{D}]^{-1}$ and its Hermitian conjugate into a geometric series, which is in fact terminated since $\bb{D}$ is a nilpotent matrix. Having known that $(D^{n})_{t,s}=\delta_{t+n\tau,s}$, the smallest integer $M$ such that $\bb{D}^M=0$ is $M=\left\lceil \frac{T}{\tau}\right\rceil$. A direct inspection of the trace gives $\Tr (\bb{D}^T)^k \bb{D}^l =(T-k\tau) \delta_{kl}$. The last term on the r.h.s. vanishes and the equation reduces to
\begin{equation}
r\alpha\sum\limits_{k=1}^{M-1}\left(\frac{\alpha^2 r^2}{|z|^2}\right)^{k-1}\frac{T-k\tau}{T-\tau}=|z|^2.
\end{equation}
Introducing $\beta=\tau/T$ and identifying $|z|$ with $s_{ext}$, one brings it to the final form
\begin{equation}
\sum_{k=1}^{M-1}\left(\frac{\alpha r}{s_{ext}}\right)^{2k}(1-k\beta)=r.
\end{equation}




\begin{thebibliography}{99}
\bibitem{WIENERMASANI}
Wiener N and  Masani P 1957  \textit{Acta Mathematica} \textbf{98}, 111-150; Wiener N and P. Masani 1958  \textit{Acta Mathematica} \textbf{99}, 93-157.
\bibitem{GRANGER}
Granger C W J 1969 \textit{Econometrica} \textbf{37}(3) 424–438.
\bibitem{MANY}
 Sims C 1980  \textit{Econometrica} \textbf{48}(1) 1–48.
\bibitem{MANY2}
Sargent T J 1987 \textit{Dynamic Macroeconomic Theory} (Harvard University Press).
\bibitem{MANY3}
 Hamilton J D 1994 \textit{Time Series Analysis}  (Princeton University Press).
\bibitem{MANY4}
L\"{u}tkepohl H 2005  \textit{New Introduction to Multiple Time Series Analysis} (Berlin: Springer).
\bibitem{MANY5}
 Hatemi-J A 2012  \textit{Empirical Economics} \textbf{43}(1) 447–456.
\bibitem{MANY6}
Berzuini C, Dawid P and Bernardinell L \textit{(ed.)} 2012 \textit{Causality: Statistical Perspectives and Applications} (Wiley).
\bibitem{WISHART}
 Wishart J 1928 \textit{Biometrika} {\bf A20} 32. 
\bibitem{WISHSINGLE}
 Silverstein J W and Choi S I 1995 \textit{J. Multivariate Anal.} \textbf{54} 295-309.
\bibitem{WISHSINGLE1}
 Burda Z, Görlich A, Jarosz A and  Jurkiewicz J 2004 \textit{Phys. A: Stat. Mech. and its Appl.} \textbf{343} 295-310.
\bibitem{WISHSINGLE2}
Vinayak and  Pandey A 2010 \textit{Phys. Rev. E} \textbf{81}  036202.
\bibitem{WISHSINGLE3}
Recher C, Kieburg M,  Guhr T and  Zirnbauer M R 2012 \textit{J. Stat. Phys.} \textbf{148} 981.
\bibitem{WISHDOUBLY}
 Waltner D,  Wirtz T and  Guhr T 2015 \textit{ J. Phys. A: Math. Theor.} \textbf{48} 175204.
 \bibitem{BURDAJURKIEWICZ}
 Burda A, Jurkiewicz J and  Wac\l aw B 2005 \textit{Phys. Rev. E} \textbf{71} 026111.
\bibitem{BOUCHAUDREVIEW}
 Bun J,  Bouchaud J-P and Potters M 2016 \textit{Phys. Rep.} \textbf{666} 1-109.
\bibitem{TELECOM}
 Couillet R and  Debbah M 2011 \textit{Random Matrix Methods for Wireless Communication} (Cambridge University Press).
\bibitem{QUANTUM}
 \.Zyczkowski K and Sommers H-J 2001 \textit{J. Phys. A: Math. Gen.} \textbf{34} 7111–7125.
\bibitem{FIN1}
 Laloux L,  Cizeau P, Bouchaud J-P and Potters M 1999 \textit{Phys. Rev. Lett.} \textbf{83} 1467. 
\bibitem{FIN2}
 Bouchaud J-P,  Potters M and  Laloux L 2005 \textit{Acta Phys. Pol. B} \textbf{36}(9) 2767–2784. 
\bibitem{GENETICS}
 Luo F \textit{et al.} 2007 \textit{BMC Bioinformatics} \textbf{8} 299.
\bibitem{GENETICS2}
 Luo F, Zhong J,  Yang Y and Zhou J 2006 \textit{Phys. Rev. E} \textbf{73} 031924.
\bibitem{METEOROLOGY}
 Preisendorfer R W 1988 \textit{Principal Component Analysis in Meteorology
and Oceanography} (New York: Elsevier).
\bibitem{ATMO}
 Santhanam M S and Patra P K 2001 \textit{Phys. Rev. E} \textbf{64} 016102.
\bibitem{CLIMATE}
Ribes A, Aza\"{i}s J-M and Planton S 2010 \textit{Clim. Dyn.} \textbf{35} 391.
\bibitem{CRIMINAL}
 Toole J L, Eagle N and Plotkin J B 2011 \textit{ACM Trans. Intell. Syst. Technol.} \textbf{2} 38.
\bibitem{EEG}
 \v{S}eba P 2003 \textit{Phys. Rev. Lett.} \textbf{91} 198104.
\bibitem{LAG1} 
 Arianos S and Carbone A 2009 \textit{J. Stat. Mech.} P03037.
\bibitem{LAG2}
 Choudhary K and Bajaj S 2012 \textit{Eurasian J. of Business and Economics} \textbf{5}(9) 165-186.
\bibitem{LIVAN}
 Livan G and Rebecchi L 2012 \textit{Eur. Phys. J. B} \textbf{85}(6) 1-11.
\bibitem{LAG3}
 Borysov S S and Balatsky A V 2014 \textit{PLoS ONE} \textbf{9}(8) e105874.
\bibitem{LAG4} 
Fiedor P 2014 \textit{Eur. Phys. J. B} \textbf{87} 168.
\bibitem{LAG5}
 Curme C,  M Tumminello,  Mantegna R N,  Stanley H E and  Kenett D Y 2015 \textit{Quant. Fin.} \textbf{15}(8)  1375-1386.
\bibitem{DROZDZ}
 Kwapie\'n J,  Dro\.zd\.z S,  G\'orski A Z and  O\'swi\k{e}cimka P 2006 \textit{Acta Phys. Pol. B} \textbf{37} 3039-3048.
\bibitem{DROZDZ2}
 Dro\.zd\.z S,  Kwapie\'n J and Ioannides A A 2011 \textit{Acta Phys. Pol. B} \textbf{42} 987-999.
\bibitem{PODOBNIK}
 Podobnik B,  Wang D,  Horvatic D,  Grosse I and  Stanley H E 2010 \textit{Eur. Phys. Lett.} {\bf 90}  68001. 
\bibitem{HINDU}
 Mayya K B K and  Amritkar R E 2006 {\it Delay Correlation Matrices}, cond-mat/0601279. 
\bibitem{USREVIEW}
 Burda Z, Jarosz A, Nowak M A, Jurkiewicz J, Papp G and  Zahed I 2011 \textit{Quant. Fin.} \textbf{11}(8) 1103-1124.
\bibitem{BOUCHAUD}
 Bouchaud J-P,  Laloux L,  Augusta M M and  Potters M 2007 \textit{Eur. Phys. Journ. B} {\bf 55} 201.
\bibitem{THURNERBIELY}
 Thurner S and Biely C 2007 \textit{Acta Phys. Pol. B} {\bf 38} 4111; Biely C and Thurner S 2008 \textit{Quantitative Finance} {\bf 8} 705-722.
\bibitem{JAROSZLAG}
 Jarosz A 2010 \textit{Hermitian and non-Hermitian covariance estimators for multivariate Gaussian and non-Gaussian assets from random matrix theory}  	arXiv:1010.2981 [q-fin.ST].
\bibitem{BJLNSproducts}
 Burda Z, Jarosz A, Livan G, Nowak M A and \'Swi\k{e}ch A 2010 \textit{Phys. Rev. E} \textbf{82} 061114.
\bibitem{BOUCHAUDLACES}
 Potters M, Bouchaud J-P and Laloux L 2005 \textit{Acta Phys. Polon.} {\bf B36} 2767.
\bibitem{EPSTEIN}
 Epstein B 1948  \textit{ Ann. Math. Statist.} \textbf{19}(3) 370.
\bibitem{VOICULESCU}
 Voiculescu D 1991 \textit{Invent. Math.} {\bf 104} 201.
\bibitem{BJNPRODUCTS}
Burda Z, Janik R and  Nowak M A 2011 \textit{Phys. Rev. E} {\bf 84} 061125.  
\bibitem{SPEICHERRAO}
 Speicher R and Rao N R 2007 \textit{Elect. Comm. in Prob.} {\bf 12} 248.    
\bibitem{HAAGERUPLARSEN}
 Haagerup U and Larsen F 2000 \textit{J. Funct. An.}  {\bf176} 331. 
 \bibitem{FEINBERGZEE}
 Feinberg J,  Scalettar R and  Zee A 2001 \textit{J. Math. Phys.} \textbf{42} 5718.
 \bibitem{HAAGLARSVECTORS}
 Belinschi S, Nowak M A,  Speicher R and  Tarnowski W, \textit{J. Phys. A: Math. Theor.} \textbf{50} 105204 (2017).
 \bibitem{BURDAJANIKWACLAW}
 Burda Z, Janik, R A and Wac\l{}aw B 2010 \textit{Phys. Rev. E} \textbf{81} 041132.
 \bibitem{SNARSKA}
 Burda Z,  Jarosz A, Nowak M A and  Snarska M 2010 \textit{New J.  Phys.} \textbf{12} 075036.
 \bibitem{SNARSKA2}
 Snarska M 2012 \textit{A Random Matrix Approach to Dynamic Factors in macroeconomic data} arXiv:1201.6544v1 [q-fin.ST].
 \bibitem{BOUCHAUDRIE}
 Bun J, Allez R, Bouchaud J-P and  Potters M 2015 \textit{Rotational invariant estimator for general
noisy matrices} arXiv:1502.06736v2 [cond-mat.stat-mech].
 \bibitem{BOUCHVECT1}
 Bun J, Bouchaud J-P and Potters M 2016 \textit{On the overlaps between eigenvectors of correlated random matrices}, arXiv:1603.04364v1 [cond-mat.stat-mech]. 
\bibitem{SOMMERS}
 Sommers H-J, Crisanti A,  Sompolinsky H and  Stein Y 1988 \textit{Phys. Rev. Lett.} \textbf{60} 1895.
\bibitem{FYODOROVSOMMERS}
 Fyodorov Y V and Sommers H-J 1997 \textit{J. Math. Phys.} \textbf{38} 1918.
\bibitem{BROWN}
 Brown L G 1983 \textit{Res. Notes Math. Ser.} \textbf{123} 1.
\bibitem{JANIKNOWAK}
 Janik R A, Nowak M A, Papp G and  Zahed I 1997 \textit{ Nucl. Phys. B}
 \textbf{501}   603–642.
 \bibitem{JARNOW}
 Jarosz A and  Nowak M A 2006 \textit{J. Phys. A} {\bf 39}  10107.
 \bibitem{CHALKERMEHLIG}
 Chalker J T and  Mehlig B 1998 \textit{Phys. Rev. Lett.} \textbf{81}  3367.
 \bibitem{CHALKERMEHLIG2}
 Chalker J T and  Mehlig B 2000 \textit{J. Math. Phys.} \textbf{41} 3233.
\bibitem{WILKINSON}
 Wilkinson J H 1965 \textit{Algebraic eigenvalue problem} (Oxford University Press).
\bibitem{TREFETHEN}
Trefethen L N and Embree M 2005 \textit{Spectra and Pseudospectra: The Behavior of Nonnormal Matrices and Operators} (Princeton University Press).
\bibitem{BERRY}
Berry M V 2003 {\it J. Modern Optics} {\bf 50} 63. 
\bibitem{SAVINSOK}
 Savin D V and  Sokolov V V 1997 {\it Phys. Rev. E} {\bf 56}  R4911.
\bibitem{FYODSAV}
 Fyodorov Y V and  Savin D V 2012 {\it Phys. Rev. Lett.} {\bf 108}  184101. 
\bibitem{FYODOROVMEHLIG}
 Fyodorov  Y V and Mehlig  B 2002 \textit{Phys. Rev. E} \textbf{66}  R045202.
\bibitem{BEENAKKER}
 Frahm K, Schomerus H,  Patra M and  Beenakker C W J 2000 {\it Europhys. Lett. } {\bf 49} 48.
\bibitem{BEENAKKER2}
 Schomerus H,  Frahm K M, Patra M and Beenakker C W J 2000 \textit{Physica A} \textbf{278}  469-496.
\bibitem{DIFFPRL}
 Burda Z,  Grela J, Nowak M A, Tarnowski W and  Warcho\l  \,P 2014
\textit{Phys. Rev. Lett.} \textbf{113} 104102. 
\bibitem{DIFFNPB}
 Burda Z, Grela J, Nowak M A, Tarnowski W and Warcho\l \, P 2015 \textit{Nucl. Phys. B} \textbf{897}  421-447.
\bibitem{JANIKNOWAKCORR}
 Janik R A, Nörenberg W, Nowak M A, Papp G and Zahed I 1999
\textit{Phys. Rev. E} \textbf{60}  2699.
\bibitem{BURDASWIECH}
 Burda Z and \'Swi\k{e}ch A 2015
\textit{Phys. Rev. E} \textbf{92} 052111.
\bibitem{WALTERSSTARR}
 Walters M and  Starr S 2015 \textit{J. Math. Phys.} \textbf{56}  013301.
\bibitem{EDELMAN}
 Edelman A, Kostlan E and Shub M 1994 \textit{J. Amer. Math. Soc.} \textbf{7} 247-267.
\bibitem{THOOFT}
 t’Hooft G 1974 \textit{Nuclear Physics B} \textbf{72} 461.
\bibitem{JAROSZGOERLICH}
 Goerlich A T and Jarosz A 2004 {\it Addition of Free Unitary Random Matrices}, arXiv:0408019 [math-ph]. 
\bibitem{JAROSZUNIT}
 Jarosz A 2011 \textit{Phys. Rev. E} {\bf 84} 011146.   
\bibitem{VDN92} 
 Voiculescu D V, Dykema K J and Nica A 1992 \textit{Free Random Variables} (Providence, RI).
\bibitem{COLLINS}
Collins B 2005 \textit{Prob. Theory Relat. Fields} {\bf 133} 315. 

 \bibitem{CHIALVO}
 Chialvo D \textit{et al.}, to appear.    

\end{thebibliography}
\end{document}